\documentclass[%
 twocolumn,
 nofootinbib,
superscriptaddress,
amsmath,amssymb,
aps,
pra,
]{revtex4-1}

\usepackage{graphicx}
\usepackage{dcolumn}
\usepackage{bm}
\usepackage{subfigure}
\usepackage{romannum}
\usepackage{physics}
\usepackage{natbib}

\usepackage[usenames,dvipsnames]{xcolor}
\definecolor{cream}{rgb}{1.0, 0.99, 0.82}
\definecolor{celadon}{rgb}{0.67, 0.88, 0.69}
\definecolor{beaublue}{rgb}{0.74, 0.83, 0.9}
\definecolor{lightorange}{rgb}{1, 0.83, 0.56}
\definecolor{lightblue}{rgb}{0.7, 0.6, 1}

\usepackage{siunitx}

\newcommand{\NV}{NV}
\newcommand{\NVm}{NV$^-$}

\begin{document}

\preprint{APS/123-QED}

\title{Amplification by stimulated emission of nitrogen vacancy centres in a diamond-loaded fibre cavity}
\author{Sarath Raman Nair}
\email{sarath.raman-nair@students.mq.edu.au, sarath91@live.com}

\author{Lachlan J. Rogers}
\affiliation{ARC Centre of Excellence for Engineered Quantum Systems (EQUS) and Department of Physics and Astronomy, Macquarie University, NSW 2109, Australia}

%
\author{ Xavier Vidal}
\affiliation{ARC Centre of Excellence for Engineered Quantum Systems (EQUS) and Department of Physics and Astronomy, Macquarie University, NSW 2109, Australia}
\affiliation{Fraunhofer Institut f{\"u}r Angewandte Festk\"orperphysik (IAF), Tullastrasse 72, 79108 Freiburg, Germany}

\author{Reece P. Roberts}
\affiliation{ARC Centre of Excellence for Engineered Quantum Systems (EQUS) and Department of Physics and Astronomy, Macquarie University, NSW 2109, Australia}

\author{Hiroshi Abe}
\affiliation{National Institutes for Quantum and Radiological Science and Technology, Takasaki, Gunma, 370-1292, Japan}

\author{Takeshi Ohshima}
\affiliation{National Institutes for Quantum and Radiological Science and Technology, Takasaki, Gunma, 370-1292, Japan}

\author{Takashi Yatsui}
\affiliation{School of Engineering, University of Tokyo, Tokyo 113-8656, Japan}

\author{Andrew D. Greentree}
\affiliation{
ARC Centre of Excellence for Nanoscale BioPhotonics, School of Science, RMIT University, Melbourne, VIC, 3001, Australia}

\author{Jan Jeske}
\affiliation{Fraunhofer Institut f{\"u}r Angewandte Festk\"orperphysik (IAF), Tullastrasse 72, 79108 Freiburg, Germany}
\affiliation{Chemical and Quantum Physics, School of Sciences, RMIT University, Melbourne 3001, Australia}

\author{Thomas Volz}
\affiliation{ARC Centre of Excellence for Engineered Quantum Systems (EQUS) and Department of Physics and Astronomy, Macquarie University, NSW 2109, Australia}


\begin{abstract}
Laser-threshold magetometry using the negatively charged nitrogen-vacancy (\NVm) centre in diamond as a gain medium has been proposed as a technique to dramatically enhance the sensitivity of room-temperature magnetometry.
We experimentally explore a diamond-loaded open tunable fibre-cavity system as a potential contender for the realization of lasing with \NVm{} centres.
We observe amplification of the transmission of a cavity-resonant seed laser at 721~nm when the cavity is pumped at 532 nm, and attribute this to stimulated emission.
Changes in the intensity of spontaneously emitted photons accompany the amplification, and a qualitative model including stimulated emission and ionisation dynamics of the \NVm{} centre captures the dynamics in the experiment very well. 
These results highlight important considerations in the realization of an \NVm{} laser in diamond.
\end{abstract}

\maketitle

\section{\label{sec:introduction}Introduction}

Quantum magnetic sensing is driving rapid development in applications ranging from life-science to navigation technologies \cite{barry2019sensitivity, degen2017quantum}, and is contributing to breakthroughs in fundamental physics \cite{tetienne2017quantum, dovzhenko2018magnetostatic, thiel2019probing}. 
Much of this effort over the last decade has made use of the negatively-charged nitrogen-vacancy (\NVm{}) centre in diamond~\cite{doherty2013nitrogen, rondin2014magnetometry, degen2017quantum, barry2019sensitivity}, which is particularly attractive since it allows quantum measurements at ambient temperatures and provides nanoscale resolution \cite{maletinsky2012robust, grinolds2013nanoscale, thiel2019probing}.
The \NVm{} centre has spin-optical properties uniquely well-suited for sensing applications~\cite{degen2017quantum}, and offers magnetometry based on reliable control and optical readout of magnetically-sensitive ground state spin levels.
Since the first theoretical proposals for \NVm spin magnetometry \cite{taylor2008high, degen2008scanning} and the breakthrough demonstrations in experiments \cite{balasubramanian2008nanoscale, maze2008nanoscale}, many different types of \NVm sensing modalities have been suggested and tested \cite{rondin2014magnetometry, barry2019sensitivity}.

While a range of experiments have been performed and even real-world technical applications are emerging, there is an ongoing quest to improve sensitivity.
The best \NVm{} sensitivity reported so far is 0.9~pT/$\sqrt{\mathrm{Hz}}$~\cite{wolf2015subpicotesla}, which is several orders of magnitude worse than initial expectations~\cite{barry2019sensitivity, degen2017quantum, taylor2008high}.
These extra orders of magnitude are required in order to reach the goal of sensitivities comparable to state-of-the-art SQUID technology \cite{barry2019sensitivity, degen2017quantum}, which would open numerous applications up to room-temperature implementation.
One of the key limiting factors is the shot-noise that arises from the difficulty in efficiently detecting single photons \cite{rondin2014magnetometry}. 
This is particularly true for single-spin sensing applications, but it still plays a role when using an ensemble of \NVm spins to enhance the sensitivity.
Ensemble magnetometry reduces the spatial resolution \cite{taylor2008high, rondin2014magnetometry, barry2019sensitivity}, but many technical applications such as magnetic resonance imaging can sacrifice spatial resolution to gain sensitivity. 

One proposal to bypass the photon shot-noise problem is a technique called laser threshold magnetometry (LTM), and it has potential to improve sensitivity towards the fT/$\sqrt{\mathrm{Hz}}$ regime~\cite{jeske2016laser, savitski2017optical}. 
The main idea is to construct a laser using the stimulated emission of the \NVm{} centres in the phonon sideband, and detect magnetic fields by the change in laser output power close to threshold.
An \NVm{} laser bypasses the shot-noise problem since the detection of spontaneous photons is replaced by the detection of a weak coherent laser beam. 
The idea of using colour centres in diamond (and in particular \NVm{} centres) as a laser gain medium has been around for a while \cite{Priv_Comm_Neil, vins2006color}.
To the best of our knowledge, the demonstration of an \NVm{} laser or even the stimulated emission of \NVm{} centres inside an optical cavity has not been achieved so far.
However, recently the first direct observation of stimulated emission from \NVm{} centres, without any optical cavity has been reported, as an important step~\cite{jeske2017stimulated}.

Stimulated emission is vital for developing a laser, and yet some contention surrounds this phenomenon in \NVm{} centres.
The main concern is the near-infrared (NIR) induced photoionisation, which has recently attracted considerable interest \cite{chen_near-infrared-enhanced_2017,hacquebard2018charge,roberts2019spin}.
Super-resolution microscopy using the stimulated emission depletion (STED) technique has been repeatedly demonstrated with \NVm{} centres \cite{rittweger2009sted, wildanger2011diffraction, wildanger2012solid, arroyo2013stimulated}, even for 1064~nm irradiation where the STED cross section should be negligible~\cite{lai_quenching_2013}.
Importantly, the STED technique will achieve superresolution with any mechanism that depletes the fluorescence of the \NVm{} centre upon irradiation of laser light, and so it does not prove the presence of stimulated emission. 
For example, photoionisation could be responsible for the quenching of fluorescence as has recently been described for low power sub-diffraction-limited microscopy using \NVm{} centres~\cite{chen_near-infrared-enhanced_2017}.
The charge-state switching dynamics between negative and neutral charge states of the NV centre due to the NIR wavelengths have also been taken into account in exploring the avenue of direct \NVm{} stimulated emission ~\cite{fraczek2017laser, hacquebard2018charge, subedi2019laser}. 
This ionization induced by NIR wavelengths naturally poses concerns about the realization of the stimulated emission from \NVm{} centres and its amplification inside a strong optical cavity for NIR wavelengths.

Here we demonstrate amplification of a laser light transmission through an open tunable fibre cavity~\cite{hunger2010fiber} loaded with a diamond sample containing high concentration of \NVm{} centres.
A 721~nm laser was used to seed the cavity mode instead of relying on the amplification of spontaneously emitted photons from the NV centres.
Optically pumping \NVm{} centres with an off-resonant \SI{532}{\nano\meter} excitation led to the amplification of this seed laser transmission.
This was accompanied by a reduction of the spontaneously emitted photons at other wavelengths, strongly supporting the conclusion of amplification by stimulated emission.
A recovery of spontaneous emission at higher pump powers is interpreted in the light of NIR induced photoionisation.
We model the observed cavity output dynamics by including charge-state switching to NV$^{0}$ and find good agreement with the data. 
Using a seed laser made it possible to observe the interplay of stimulated emission and charge state dynamics in the presence of a cavity with lower requirements on the effective cavity quality factor.
These results address the concern that the strong NIR intra-cavity field might induce charge state switching of NV$^{-}$ to NV$^{0}$ to a degree that precludes an \NVm{} laser.
Our analysis highlights the important parameters, and processes occurring in the diamond samples which are currently limiting the realisation of an NV laser threshold magnetometer as originally proposed in Ref~\cite{jeske2016laser}.

This article has the following structure.
The diamond sample and experimental fibre-cavity apparatus are introduced in section \ref{sec:exp_details}.
Section \ref{sec:exp_results} presents the cavity transmission measurements showing optical amplification.
A rate equation model is developed in Section \ref{sec:model} that demonstrates how stimulated emission and charge state switching of \NV{} centres can account for the observed amplification behaviour.
The concluding Section \ref{sec:conclusion} discusses the challenges ahead for achieving NV$^{-}$ lasing.

%
\section{\label{sec:exp_details}Experimental details}

\begin{figure*}
\centering
\subfigure[]
{\label{fig:cavity_illustration}\includegraphics[scale = 1]{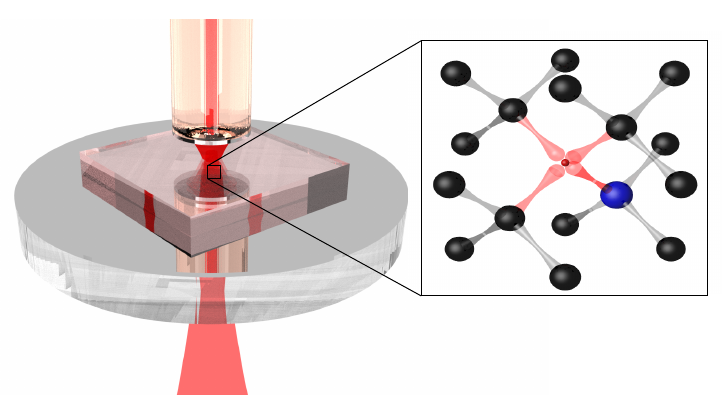}}
\subfigure[]
{\label{fig:cavity_modes}\includegraphics[scale = 0.5]{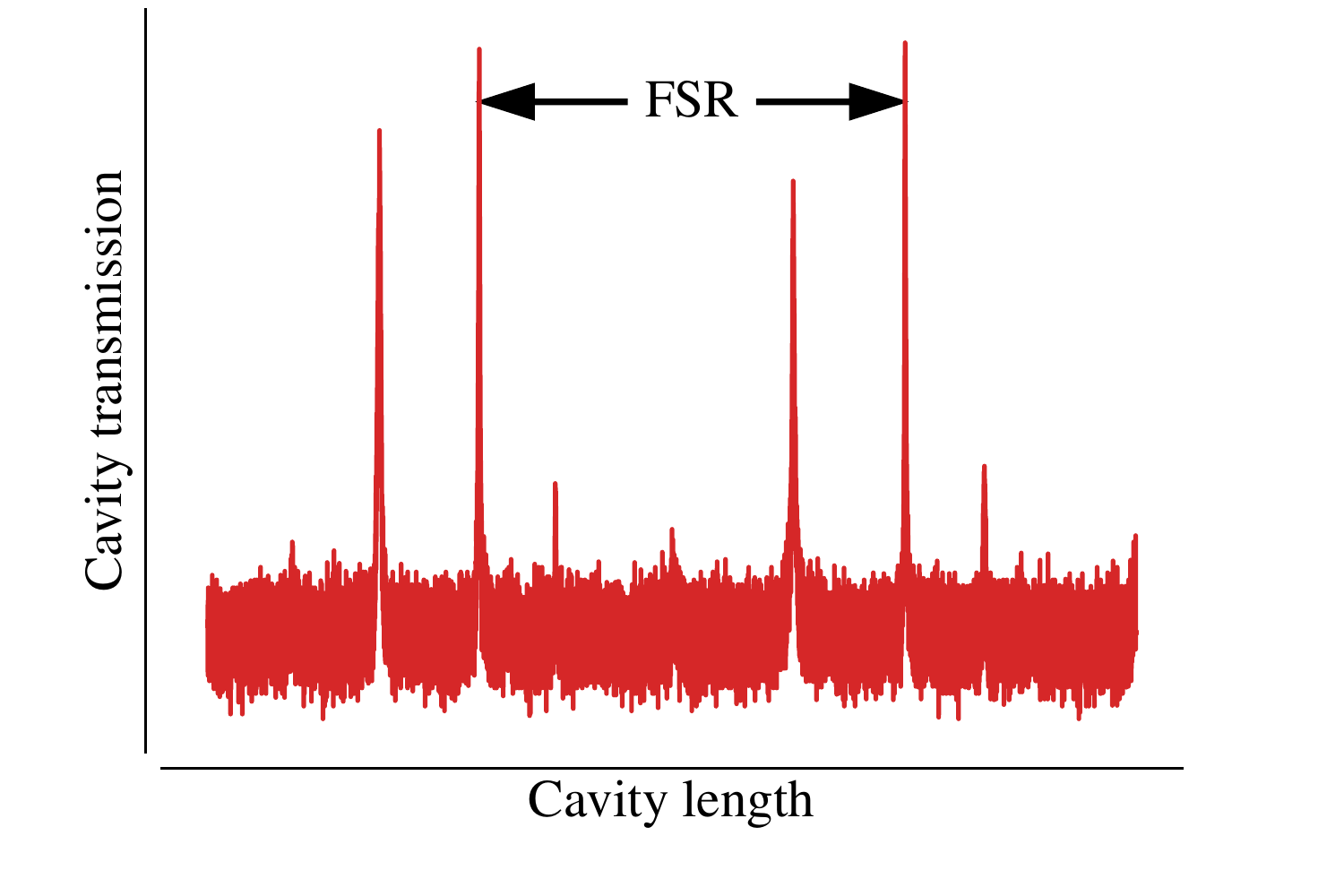}}
\caption{
    Diamond loaded fibre cavity platform 
    (a) Schematic of the fibre-cavity loaded with a diamond sample containing \NV\,centres. 
    The diamond sample is placed on top of the macroscopic mirror and a concave fibre-end mirror is approached from the top forming a stable Fabry-Perot microcavity.
    The inset displays the structure of the \NV\,centre within the diamond lattice.
    Black spheres are carbon atoms, blue indicates nitrogen and red displays electron orbitals around the vacancy position.
    (b) An illustration of cavity modes recorded for an arbitrary wavelength in the high reflective stop band of the mirror coatings. The free-spectral range (FSR) is annotated in the figure.
}
\label{fig:1}
\end{figure*}

The sample used in these experiments was a commercial type 1b HPHT diamond from E6 with [100] orientation, which was electron-beam irradiated at \SI{2}{\mega\electronvolt} to a fluence of $10^{18}$cm$^{-2}$. 
This was performed in vacuum and at a temperature of \SI{740}{\celsius} in order to achieve in-situ annealing, which improves NV creation \cite{Capelli2018} to a concentration of \SI{1.7}{ppm}.

This corresponds to an effective density of NV centres of \SI{3e17}{\per\cm\cubed}, which is the number used in our models for the remainder of the paper.
The sample was then laser cut and polished to a thickness of \SI{50}{\micro\metre} and afterwards optically ``superpolished'' with a pulsed laser at \SI{213}{\nano\metre} and \SI{24}{\micro\joule} pulse power under oxygen atmosphere. 
This achieved a very high surface smoothness of $R_\mathrm{a}$=\SI{0.4}{\nano\metre}, in an attempt to eliminate scattering losses from the surface.

In order to investigate \NVm{} lasing, this diamond was placed inside an open, mechanically tunable fibre cavity \cite{hunger2010fiber}, consisting of a concave fibre-end mirror and a flat mirror substrate, forming a stable semi-concave cavity as illustrated in Figure \ref{fig:1}.
The curvature radius of the fibre-tip mirror was around \SI{170}{\micro\metre}.
This imposes an upper limit on the stable cavity length, but the high refractive index of diamond means that placing the sample in the cavity increases the effective maximum cavity length.
This ensures that a stable cavity mode can be formed while maintaining a safe separation between the fibre tip and the diamond sample.
The fibre was mounted on a shear piezo (Noliac NAC2402-H3.4) to allow fine and rapid control of the cavity length.
Both cavity mirrors were coated with the same distributed Bragg reflector (DBR) coating (Laseroptik GmbH, Germany) designed to have two stopbands, one at \SI{633}{\nano\metre}, the other one at \SI{740}{\nano\metre}. 
The cavity was mechanically designed to achieve passive stability sufficient for this study.

Green off-resonant pump light at \SI{532}{\nano\metre} (Lighthouse Photonics) and seed red light at \SI{721}{\nano\metre} (MSquared SOLSTIS) were both delivered into the fibre cavity through the fibre-mirror end. 
At \SI{532}{\nano\metre} the dielectric DBR coating transmits roughly 60$\%$ of the incoming light generating a poor cavity mode.
This makes it possible to achieve high pump laser powers at the sample, regardless of cavity tuning.  
A red laser wavelength at \SI{721}{\nano\metre} was chosen over a \SI{740}{\nano\metre} laser where the cavity reflectively is at a maximum because a lower reflectively will produce a lower finesse cavity which is less susceptible to noise and easier to handle without electronic stabilization.
The designed finesse for the cavity around 721nm is around 40000.
Experimentally measured cavity finesse was 10--100 times lower, limited by many factors such as light absorption in and surface scattering off the diamond sample. 
The light exiting the cavity through the macroscopic mirror substrate was collected and sent for analysis to a spectrometer (ACTON/Princeton Instruments SpectraPro 2500i).
Before detection, the green pump light was blocked with a \SI{532}{\nano\metre} notch filter.
In the absence of an active feedback mechanism, we captured the cavity resonance for the \SI{721}{\nano\metre} laser by applying a triangle-ramped voltage to the piezo that controls the cavity length. 
\begin{figure}
\centering
{\includegraphics[width = 7cm, height = 5cm]{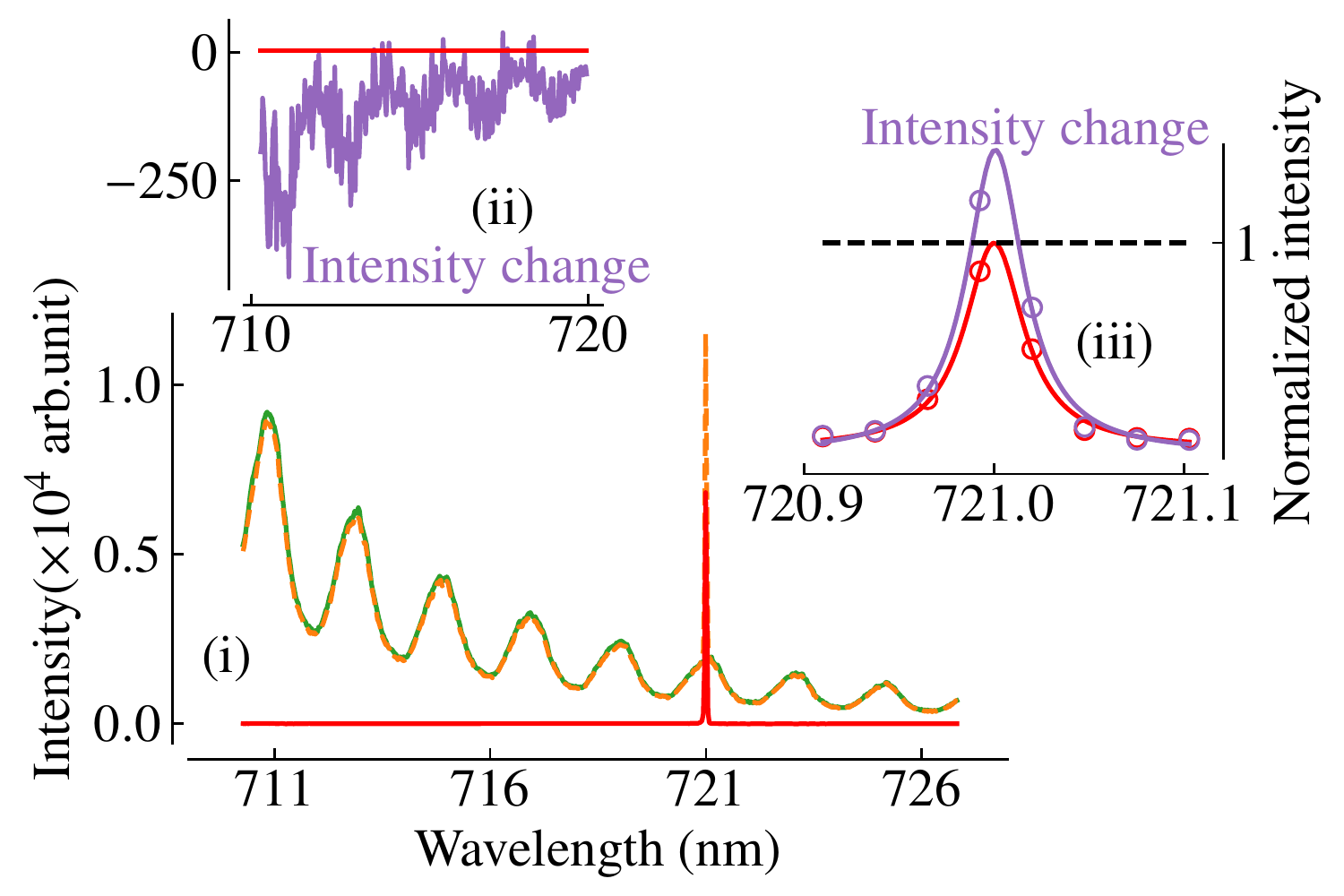}}
\caption{
    Typical experimental spectra obtained in cavity transmission. 
    The green, orange and red traces in (i) were obtained with green laser only, with green plus red laser and with red laser only present, respectively. 
    The modulation pattern with a period of about 2.11~nm in (i) is due to the weak cavity modes formed inside the diamond crystal itself.
    Subtracting the green trace from the orange trace yields the purple trace in (ii) and the data points shown in (iii) as purple circles (annotated as intensity change in both figures). 
    The red trace in (ii) is the spectrum with only the red laser near $721$~nm present shown in (i). 
    The red and purple lines shown in (iii) is the Lorentzian fit to the data points shown in respective colours.
    In the trace with green plus red laser present, a slight decrease of the spontaneous emission occurs compared to the green laser only trace, which is apparent in (ii).
    Amplification of the red laser is apparent at the cavity mode resonant with the red laser in (iii). 
}
\label{fig:2}
\end{figure} 
The spectrometer camera integration time was set such that all photons arriving at the detector during a single cavity ramp are captured.
Spectrometer camera and cavity ramp were time-synchronized.

\begin{figure}
\centering
\includegraphics[scale = 0.5]{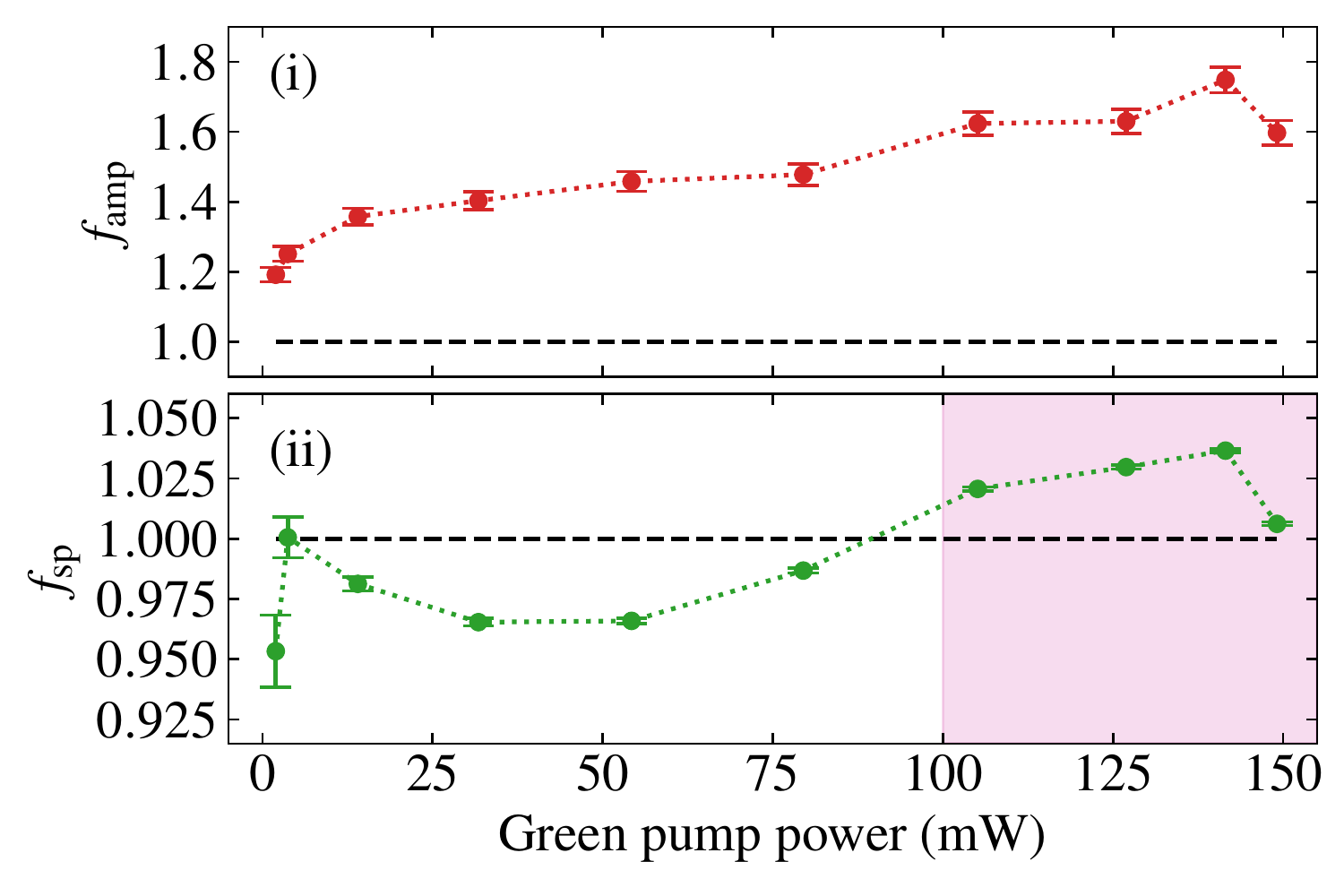}
\caption{Amplification and spontaneous-emission factor as a function of green pump power. Error bars shown are one standard deviation range obtained by propagating the shot noise in spectrometer counts for each pump powers (\romannum{1}) $f_\mathrm{amp}$ was extracted from raw data like the ones displayed in Figure~\ref{fig:2}\, 
 and measures the amplification of the red laser upon off-resonant pumping of NV centres. (\romannum{2}) $f_\mathrm{sp}$ measure the influence of the red laser on the spontaneous emission of the NV centres and was obtained by integration of the background signals in the wavelength region of the inset of Figure~\ref{fig:2}~(ii). The black dotted lines indicates the reference line at unity, which corresponds to no difference in output signal between the different pumping schemes.}
\label{fig:3}
\end{figure}
\begin{figure*}
\centering
\subfigure[]
{\label{fig:4.1}\includegraphics[scale = 0.5]{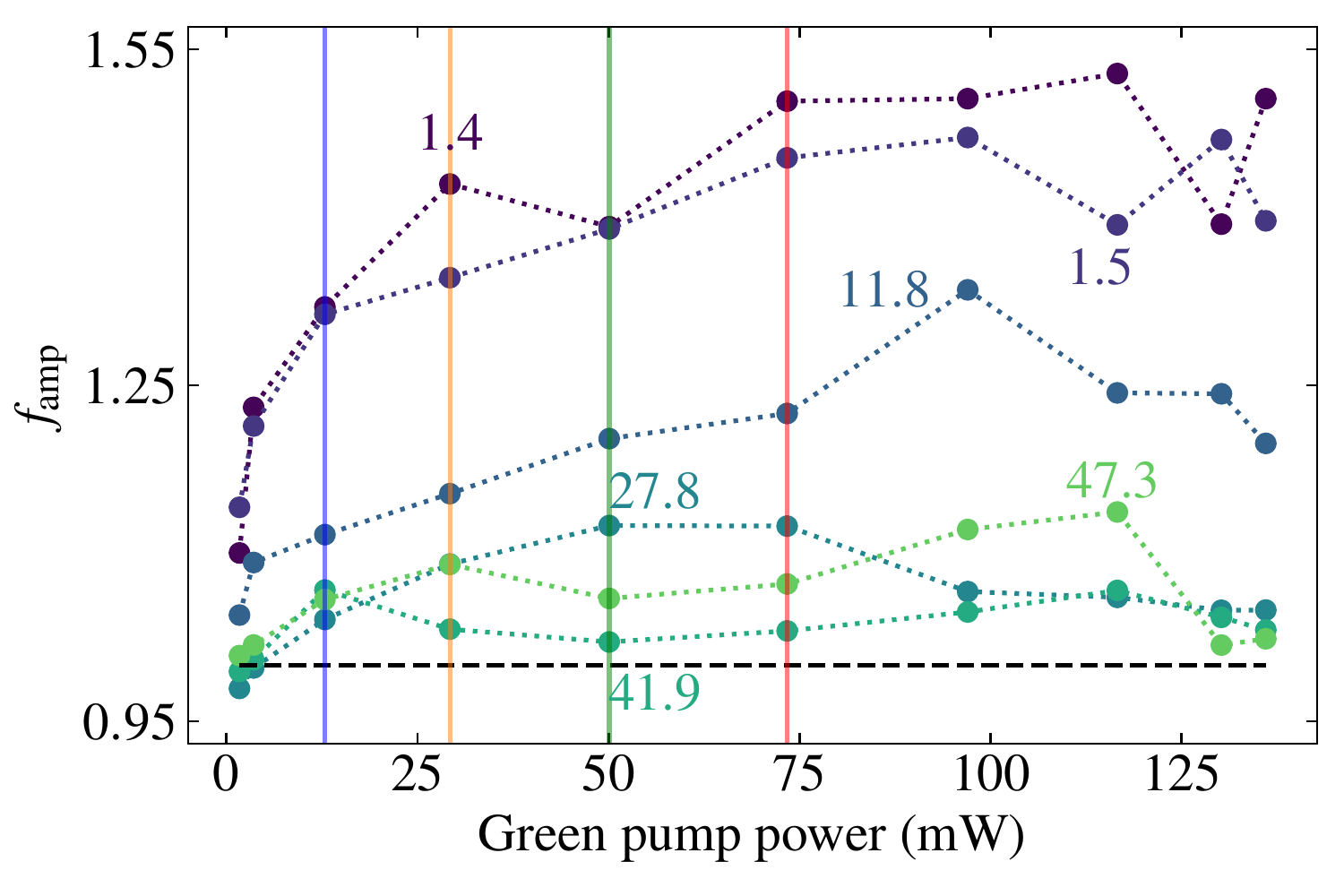}}
\subfigure[]
{\label{fig:4.2}\includegraphics[scale = 0.5]{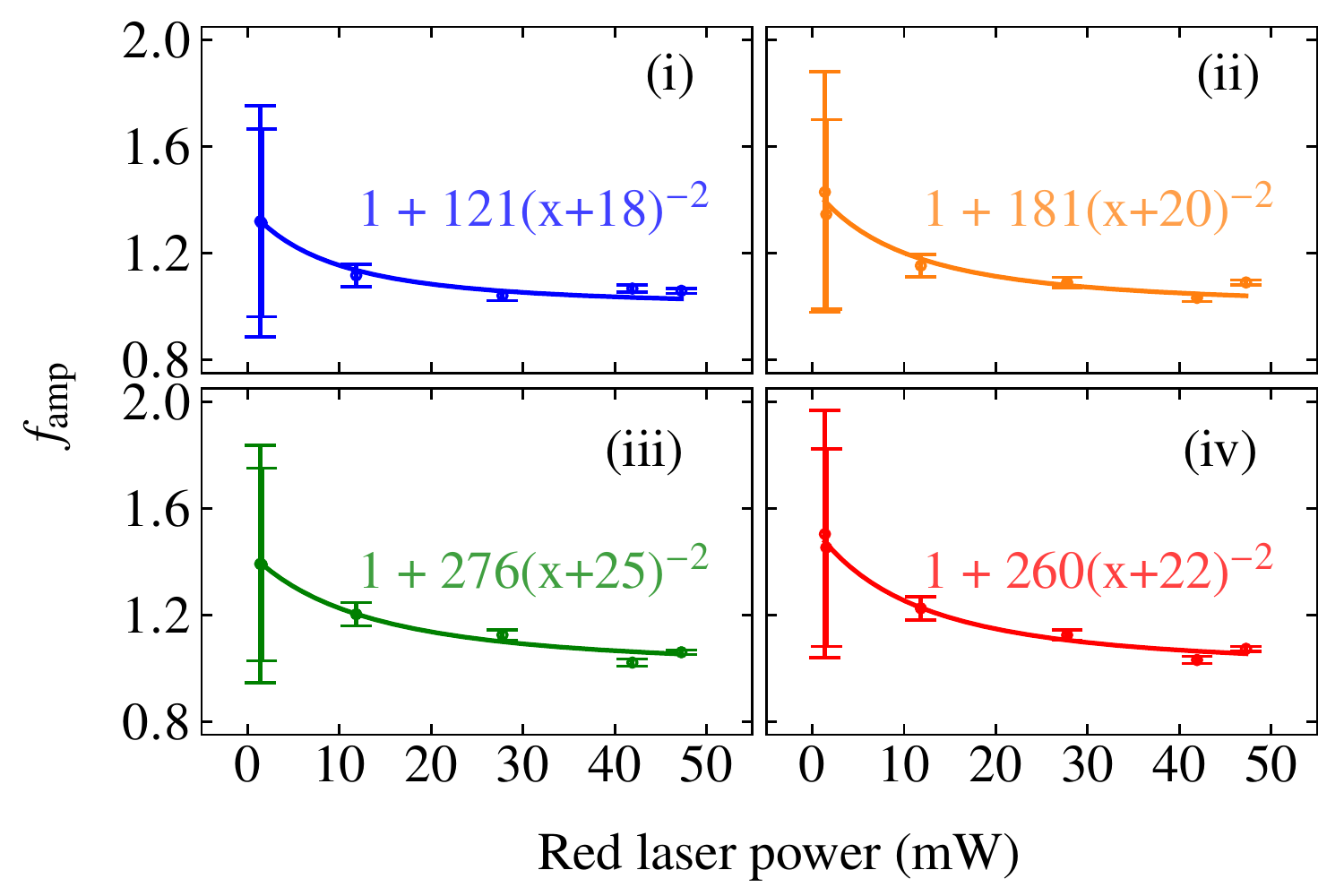}}
\caption{Dependence of $f_\mathrm{amp}$ on red power. (a) $f_\mathrm{amp}$ as a function of green pump power for different red powers. Red laser power in mW, corresponding to each curve is annotated in respective colours. 
(b) Line-cuts through (a) give $f_\mathrm{amp}$ as a function of red laser power for fixed green pump power. The blue, orange, green, and red circles in (\romannum{1}), (\romannum{2}), (\romannum{3}), and (\romannum{4}) correspond to green pump powers of $\sim$ 12.9 mW, $\sim$ 29.2 mW, $\sim$ 50.1 mW, and $\sim$ 73.4 mW respectively. The solid curves with respective colors are from the  fit functions annotated in each figures. For clarity, error bars, which are obtained in the same manner as Figure \ref{fig:3}, are shown only in (b)}
\label{fig:4}
\end{figure*}

\section{\label{sec:exp_results} Experimental Results}

Two typical spectra under different pumping conditions are displayed in Figure~\ref{fig:2}. 
The solid green trace was recorded for \SI{532}{\nano\meter} off-resonant green pumping alone, whereas the dashed orange trace was recorded for the same green pumping intensity with the red laser added at \SI{721}{\nano\meter}.
In essence, both spectra show a broad spontaneous emission background from the NV centres phonon sideband that is modulated by the transmission of the mirror coatings and interference effects caused by the \SI{50}{\micro\meter} diamond sample.  
The broad modulation feature with a period of around 2.11~nm is due to the diamond sample, and the period matches what is expected from the thickness of the diamond sample.
Clearly, the spectra have very similar broad backgrounds but the orange trace with the red laser added shows the \SI{721}{\nano\meter} resonant cavity mode appearing above the NV phonon side band.

In order to see the difference between the two traces, we subtract the orange trace (red plus green lasers) from the green trace (green laser alone).
Comparing the resulting trace with a spectra of the red \SI{721}{\nano\meter} laser alone, shown in red on the same plot, allows us to observe the impact of the additional red laser on NV centres excited by the green laser.
The first impact of the red laser on the excited NV centres is a slight reduction in the background fluorescence (Figure~\ref{fig:2}(ii)). 
Secondly, the red laser peak has an increased intensity indicating an amplification of the power in the \SI{721}{\nano\meter} cavity mode (Figure~\ref{fig:2}(iii)).
As expected, the red trace is flat in the range between \SI{711}{\nano\meter} and \SI{721}{\nano\meter} of the spectrum since under red pumping alone no noticeable spontaneously emission photons from \NV{} centres are detected. 
In contrast to the amplified laser peak, the purple trace shows a negative background with a clear modulation contrast reminiscent of the large modulated background in part (i). 
This indicates that when both lasers are present, the spontaneous background is suppressed whilst the red laser being amplified. 

We vary the excitation power and observe the output of the cavity mode in a systematic fashion.
In a first set of measurements, we keep the red power around 67~$\mu$W and we vary the green pump power roughly between 1~mW and 150~mW. Both numbers refer to the input power into the cavity fibre. 
At each green pump power we obtain a series of spectra as in Figure~\ref{fig:2} (i).
We then compare the corresponding purple and red traces (Figure~\ref{fig:2} (iii)) to extract the amplification factor for the resonant laser in a quantitative way.
The dimensionless ratio of the two, $f_\mathrm{amp}$, obtained by the numerical summation of the data points, is a measure of the amplification of the red laser light transmitting through the cavity under green pumping.
Figure~\ref{fig:3}~(i) displays the resulting amplification factor  $f_\mathrm{amp}$ as a function of green pump power. 
Clearly, we see an amplification factor above 1 that monotonically increase over the power range plotted here (except the last data point). 
However, the trend shows a strong saturation behavior and the further increase in amplification for higher green pump power is only weak.
In addition, we extract a quantitative measure for the effect of the red laser on the spontaneous emission from the NV centres.
We therefore take the orange curve in Figure~\ref{fig:2} (green plus red laser) and subtract the red curve (red laser only) from it.   
For the resulting trace (not displayed in Figure~\ref{fig:2}), we numerically sum the data points for the range shown in Figure~\ref{fig:2}(ii).  
The ratio $f_\mathrm{sp}$ of this integrated intensity to the integrated intensity of the corresponding wavelength region in the green spectrum of Figure~\ref{fig:2} (i) (green laser only), gives a quantitative measure for the effect of the red laser on the NV spontaneous emission.
Interestingly, when plotting $f_\mathrm{sp}$ as a function of green pump power (Figure~\ref{fig:3}~(ii)), we observe an initial decrease in spontaneous emission as expected in the presence of stimulated emission.  
But as we increase the green pump power beyond \SI{50}{\milli\watt}, the trend reverses and somewhat surprisingly, we see more NV centre spontaneous emission in the presence of the resonant laser than without it (shaded magenta region in \ref{fig:3}~(ii)). 
Interestingly, $f_\mathrm{sp}$ seems to follow the behavior of the corresponding $f_\mathrm{amp}$ in the shaded region.
Both the strong saturation of the amplification factor, $f_\mathrm{amp}$, and the `reversing' behaviour in the spontaneous emission factor, $f_\mathrm{sp}$, raise questions about the underlying physical mechanism.
In the following section, we therefore present a qualitative model that sheds some light on the physics going on in our diamond-cavity system section. 

This measurement of amplification as a function of green pump power was repeated for a systematic range of the red probe laser powers.
The observed effect of the red laser power on $f_\mathrm{amp}$ is summarized in Figure~\ref{fig:4}. 
As the red power increases from around 1~mW (blue) to around 47~mW (purple), amplification gets increasingly suppressed, as witnessed by the fact that $f_\mathrm{amp}$ saturates much more rapidly and essentially stays flat around ~1 for the highest red powers. 
In order to quantify this effect, in Figure~\ref{fig:4.2} we plot $f_\mathrm{amp}$ as a function of red power for fixed green pump powers (corresponding to the marked vertical slices through the data in Figure~\ref{fig:4.1}).
These data fit well with an inverse square law as demonstrated by the fit curves displayed alongside the data points.

Finally, we would like to point out that in different iterations of our experiment (with slight changes to setup and cavity parameters), we have consistently observed amplification factors up to a maximum of $\sim$3, and comparable reversing trends in the spontaneous emission signal that also include the peculiar increase in $f_\mathrm{sp}$ above 1 for higher green pump powers. 
In addition to this we have observed reduction in the amplification factor $f_\mathrm{amp}$ below 1 in some of our measurements with high red laser power.
This is also apparent in some of the data points, particularly for low green pump powers and high red laser power in figure \ref{fig:4.1}.

\section{\label{sec:model} A Qualitative model based on stimulated emission and charge state switching}
\begin{figure*}
\centering
\includegraphics[scale = 0.6]{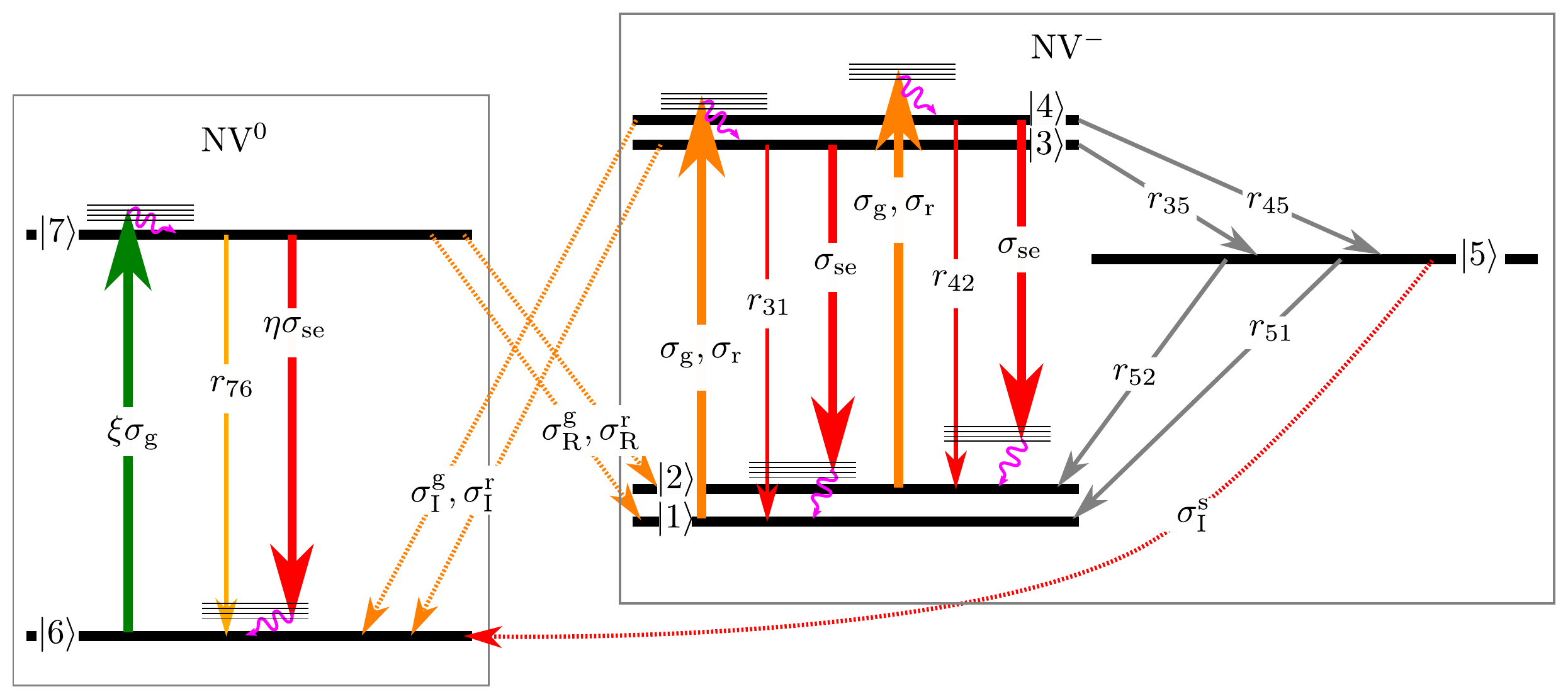}
\caption{NV centre model for explaining amplification. 
Levels 1 to 5 belong to NV$^{-}$ and 6 and 7 belong to NV$^{0}$.
In the NV$^{-}$, 1 and 2 respectively represent the spins 0 and 1 of the ground state, and 3 and 4 respectively represent the spins 0 and 1 of the excited state.
5 represent the singlet level of the NV$^{-}$ centre.
In the case of NV$^{0}$, it has only ground state (6) and excited state (7).
The orange arrows represent transitions due to both green and red wavelengths. 
The solid orange line represents the pumping rate, whereas the dotted one represent the ionization and recombination transitions.
The thin and thick solid red arrows represent spontaneous and stimulated transitions.
The grey arrows represent inter-system crossing (ISC) transitions and curly magenta arrows represent phonon decays.
The green arrow represent the pumping rate of NV$^{0}$ by the green wavelength.
The transition rates of each transition is shown on each corresponding arrows.
}
\label{fig:5}
\end{figure*}

In this section, we theoretically model an ideal amplification experiment taking into account the photo-induced charge inter-conversion between NV$^{-}$ and NV$^{0}$, as well as the stimulated emission from these charge states.
The results from the theoretical modelling allow us to qualitatively interpret the experimental results of the previous section.

\subsection{Model}

We consider an ideal cavity on resonance with the red laser at \SI{721}{\nano\metre}  in the fundamental (Gaussian) mode.
For ease of calculation, we assume a flat-top approximation and a cylindrical volume of the cavity with the radius of the Gaussian spot size around \SI{5}{\micro\meter} on the macroscopic mirror as the base radius.
We further assume that the cavity is fully filled with the diamond sample.
The effective cavity volume determines the number of NV centres contributing to the dynamics and the laser amplification.
It further sets the stimulating intra-cavity power/intensity of the red laser by scaling the power that we provide at the fibre input by the effective finesse of the cavity.

In order to model the NV centres in our cavity, we first consider a single NV centre and capture its internal dynamics including charge state switching. 
We then simply scale the population fractions with the total number of NV centres inside the cavity volume to obtain the ensemble effect.
A review of the NV centre charge state switching dynamics is given in Appendix~\ref{sec:addendixSwitching} and the resulting level scheme and transitions are depicted in Figure~\ref{fig:5}.
We can then write down the rate equations based on the level scheme in Figure~\ref{fig:5}.
\begin{widetext}
\begin{align}
\label{eq:3}
\dv{p_{\ket{1}}}{t} =& -\left(\frac{I_g\sigma_g}{h\nu_g}+\frac{I_r\sigma_r}{h\nu_r}\right)p_{\ket{1}}+r_{31}p_{\ket{3}}+\frac{I_r\sigma_{se}}{h\nu_r}p_{\ket{3}}+r_{51}p_{\ket{5}}+\frac{1}{2}\left(\frac{I_g\sigma^g_R}{h\nu_g}+\frac{I_r\sigma^r_R}{h\nu_r}\right)p_{\ket{7}},\\
\label{eq:4}
\dv{p_{\ket{2}}}{t} =& -\left(\frac{I_g\sigma_g}{h\nu_g}+\frac{I_r\sigma_r}{h\nu_r}\right)p_{\ket{2}}+r_{42}p_{\ket{4}}+\frac{I_r\sigma_{se}}{h\nu_r}p_{\ket{4}}+r_{52}p_{\ket{5}}+\frac{1}{2}\left(\frac{I_g\sigma^g_R}{h\nu_g}+\frac{I_r\sigma^r_R}{h\nu_r}\right)p_{\ket{7}},\\
\label{eq:5}
\dv{p_{\ket{3}}}{t}=& -\left(r_{31}+r_{35}+\frac{I_g\sigma^g_I}{h\nu_g}+\frac{I_r\sigma^r_I}{h\nu_r}\right)p_{\ket{3}}+\left(\frac{I_g\sigma_g}{h\nu_g}+\frac{I_r\sigma_r}{h\nu_r}\right)p_{\ket{1}},\\
\label{eq:6}
\dv{p_{\ket{4}}}{t}=& -\left(r_{41}+r_{45}+\frac{I_g\sigma^g_I}{h\nu_g}+\frac{I_r\sigma^r_I}{h\nu_r}\right)p_{\ket{4}}+\left(\frac{I_g\sigma_g}{h\nu_g}+\frac{I_r\sigma_r}{h\nu_r}\right)p_{\ket{2}},\\
\label{eq:7}
\dv{p_{\ket{5}}}{t}=&-(r_{52}+r_{51}+\frac{I_r\sigma^s_I}{h\nu_r})p_{\ket{5}}+r_{35}p_{\ket{3}}+r_{45}p_{\ket{4}},\\
\label{eq:8}
\dv{p_{\ket{6}}}{t}=&-\frac{I_g(\xi\sigma_g)}{h\nu_g}+\left(r_{76}+\frac{I_r(\eta\sigma_{se})}{h\nu_r}\right)p_{\ket{7}}+\frac{I_r\sigma^s_I}{h\nu_r}p_{\ket{5}}+ \left(\frac{I_g\sigma^g_I}{h\nu_g}+\frac{I_r(\sigma^r_I)}{h\nu_r}\right)\left(p_{\ket{3}}+p_{\ket{4}}\right),\\
\label{eq:9}
\dv{p_{\ket{7}}}{t}=&-(r_{76}+\frac{I_r(\eta\sigma_{se})}{h\nu_r}p_{\ket{7}}+\left(\frac{I_g\sigma^g_R}{h\nu_g}+\frac{I_r\sigma^r_R}{h\nu_r}\right)+\frac{I_g(\xi\sigma_g)}{h\nu_g}p_{\ket{6}}.
\end{align}
\end{widetext}
Here $p_{\ket{j}}$ is the population fraction of a single NV centre on the level $\ket{j}$.
Thus $\sum_{j=1}^{7}p_{\ket{j}} = 1$.
Since the decay from the phonon levels of the ground state is expected to be very high, we neglect any stimulated absorption from these phonon levels back to the excited states.
From the model, we obtain the steady-state population fractions of each of the seven levels, denoted by $p_{i}$ with $i=1,...,7$.
The spontaneous and stimulated emission intensity of the NV centres can then be calculated from the steady state populations in the excited states of both the NV$^-$ charge state ($p_{\ket{-}}$ = $p_{\ket{3}}$ + $p_{\ket{4}}$) and the NV$^0$ charge state ($p_{\ket{0}}$ = $p_{\ket{7}}$).
In our qualitative model we make the following assumptions for simplicity.
Firstly, whilst the coupling between each NV centre in the diamond and the cavity mode really is different~\cite{janitz2015fabry}, we assume every NV centre in the diamond to have the same coupling for simplicity of calculation and in order to extract the essential physics.
This model also neglects possible dipole-dipole interactions and coherent collective effects~\cite{bradac_room-temperature_2017}, which is reasonable given the typical cavity decay rates and the NV density in our system.
In addition, since we are using high intensity fields we neglect the single photon ionisation process that have been observed for low power off-resonant excitation~\cite{manson_nv_2018}. 
We also do not consider the spin dependence on ionisation transitions directly from excited states that has recently been reported~\cite{roberts2019spin} as it is not necessary in order to capture the important physics and trends observed in the data. 
Furthermore, emission of the NV$^0$ which may then be subsequently captured by surrounding NV$^-$ centres has not been considered in this model. 
In dense NV samples this effect will reduce the effective contribution of NV$^0$ in the emitted signal, however, we expect this effect to be small.
Lastly, we do not consider any change in the cavity conditions that modify the input light into the cavity and there by change the intra-cavity field as well as the cavity transmission for simplicity.

Both charge states can contribute to the stimulated emission signal since both charge states have a broad phonon sideband with significant amplitude at 721~nm.
The amplification factor, $f_\mathrm{amp}$, which is a ratio of total intra-cavity power at \SI{721}{\nano\metre}, to the intra-cavity power of the seed laser.
Assuming stimulated emission of the two charge states of NV centres as the additional contribution to the intra-cavity power at \SI{721}{\nano\metre}, other than the external seed laser at \SI{721}{\nano\metre}, the $f_\mathrm{amp}$ can be written as,
\begin{equation}
\label{eq:1}
f_{\mathrm{amp}} = 1 + \frac{1}{2} \Bigl((p_{\ket{-}} + \eta p_{\ket{0}}) F \sigma_{\mathrm{se}} \rho_{\mathrm{NV}} l\Bigr) .
\end{equation}
Here $\sigma_{se}$ is the stimulated emission cross-section of the NV centre in the negative charge state. 
$\eta$ represents the ratio of stimulated emission cross-sections of NV$^{0}$ to NV$^{-}$.
$\rho_{NV}$ is the density of the NV centres in the diamond sample and $l$ is the cavity length, which in our simplified model we assume to be the thickness of the sample. 
$F$ denotes the ratio of the intra-cavity resonant (red) laser power and the input power at the fibre input, which can be approximated as finesse/$\pi$.
The factor $\frac{1}{2}$ reflects the fact that we observe the stimulated emission only through the bottom mirror and therefore capture only half of the total stimulated power.
It is important to note that the assumption leading to this equation's simplicity is that the change in spontaneous emission when adding the red seed laser is small compared the stimulated emission cross section, which is valid for the seed intensities used in this paper.

Next, we can obtain the spontaneous emission factor, $f_\mathrm{sp}$, by comparing the number of photons spontaneously emitted under green plus red pumping compared to green pumping only. 
We include both charge states and write $f_\mathrm{sp}$ as
\begin{equation}
\label{eq:2}
f_{\mathrm{sp}} = \frac{p_{\ket{-}}+\beta p_{\ket{0}}}{p^{'}_{\ket{-}}+\beta p^{'}_{\ket{0}}}.
\end{equation}
The prime $'$ marks the population fractions under green pumping only. 
$\beta$ is the ratio of the spontaneous emission rate of NV$^{0}$ to NV$^{-}$. 
We neglect the emission from the NV$^{0}$, when there is only red laser present, since the red laser cannot excite this centre in our model.

\subsection{\label{sec:Numerical} Numerical results}

To model the NV centre, we have approximated each of the parameters by obtaining values from the literature, using values from HPHT type 1b bulk diamond where possible. These parameters are shown in Table~\ref{tab:1}.

\begin{table}
\centering
\begin{tabular}{llc}
\hline
\textbf{Parameter} & \textbf{Value} & \textbf{Reference}\\
\hline
$r_{31} = r_{42}$  & 63.93 MHz & \cite{tetienne2012magnetic} \\
$r_{35}$ & 7.93 MHz & \cite{tetienne2012magnetic}\\
$r_{45}$ & 53.25 MHz & \cite{tetienne2012magnetic}\\
$r_{51}$ & 0.98 MHz & \cite{tetienne2012magnetic}\\
$r_{52}$ & 0.72 MHz & \cite{tetienne2012magnetic}\\
$r_{76}$ & 0.74 $\times$ $r_{31}$ & \cite{hacquebard2018charge}\\
$\sigma_{se}$ & 3.22 $\times$ 10$^{-21}$m$^2$ & --\\
$\sigma_{g}$ & 3.1 $\times$ 10$^{-21}$m$^{2}$ & \cite{wee2007two}\\
$\sigma_{r}$ & 3 $\times$ 10$^{-24}$m$^{2}$ & \cite{jeske2017stimulated}\\
$\sigma_{I}^{g}$ & 0.037 $\times$ $\sigma_{g}$ & \cite{hacquebard2018charge}\\
$\sigma_{R}^{g}$ & 0.08 $\times$ $\sigma_{g}$ & \cite{hacquebard2018charge}\\
$\sigma_{I}^{r}$ & 0.071 $\times$ $\sigma_{se}$ & \cite{hacquebard2018charge}\\
$\sigma_{R}^{r}$ & 0.22 $\times$ $\sigma_{se}$ & \cite{hacquebard2018charge}\\
$\sigma^s_{I}$ & 0.0215 $\times$ $\sigma_{se}$ & \cite{hacquebard2018charge}\\

\hline
\end{tabular}
\caption{Parameters used for the theoretical estimations. The transition rates are the average of the values from ref \cite{tetienne2012magnetic}.
$\sigma_{g}$ and $\sigma_{r}$ are absorption cross-sections of NV$^{-}$ for green and red wavelengths.
The absorption cross-section of the NV$^{0}$ for green wavelength is $\xi=1.3$ times higher than that of NV$^{-}$ \cite{hacquebard2018charge}.
Though the wavelength used in \cite{jeske2017stimulated} is different, absorption cross-section for the stimulating wavelength is considered approximately equal to this value.}

\label{tab:1}
\end{table}

The stimulated emission cross-section $\sigma_{se} \sim 3.22\times10^{-21}$m$^{2}$ for our sample is obtained in a similar fashion to reference \cite{fraczek2017laser}, but considering only the NV$^{-}$ centres (see appendix \ref{sec:appendix2}).
The value we obtain for our system is comparable to the values obtained in literature previously~\cite{fraczek2017laser, subedi2019laser, vins2006color}.
Near our working wavelength of $721$~nm, reference~\cite{fraczek2017laser} specifies a ratio between the stimulated emission cross-section of NV$^{-}$ and NV$^{0}$ of slightly higher than 3.
We hence assume $\eta_{se} = \frac{1}{3}$ in our model.

\begin{figure}
\centering
\includegraphics[scale = 0.5]{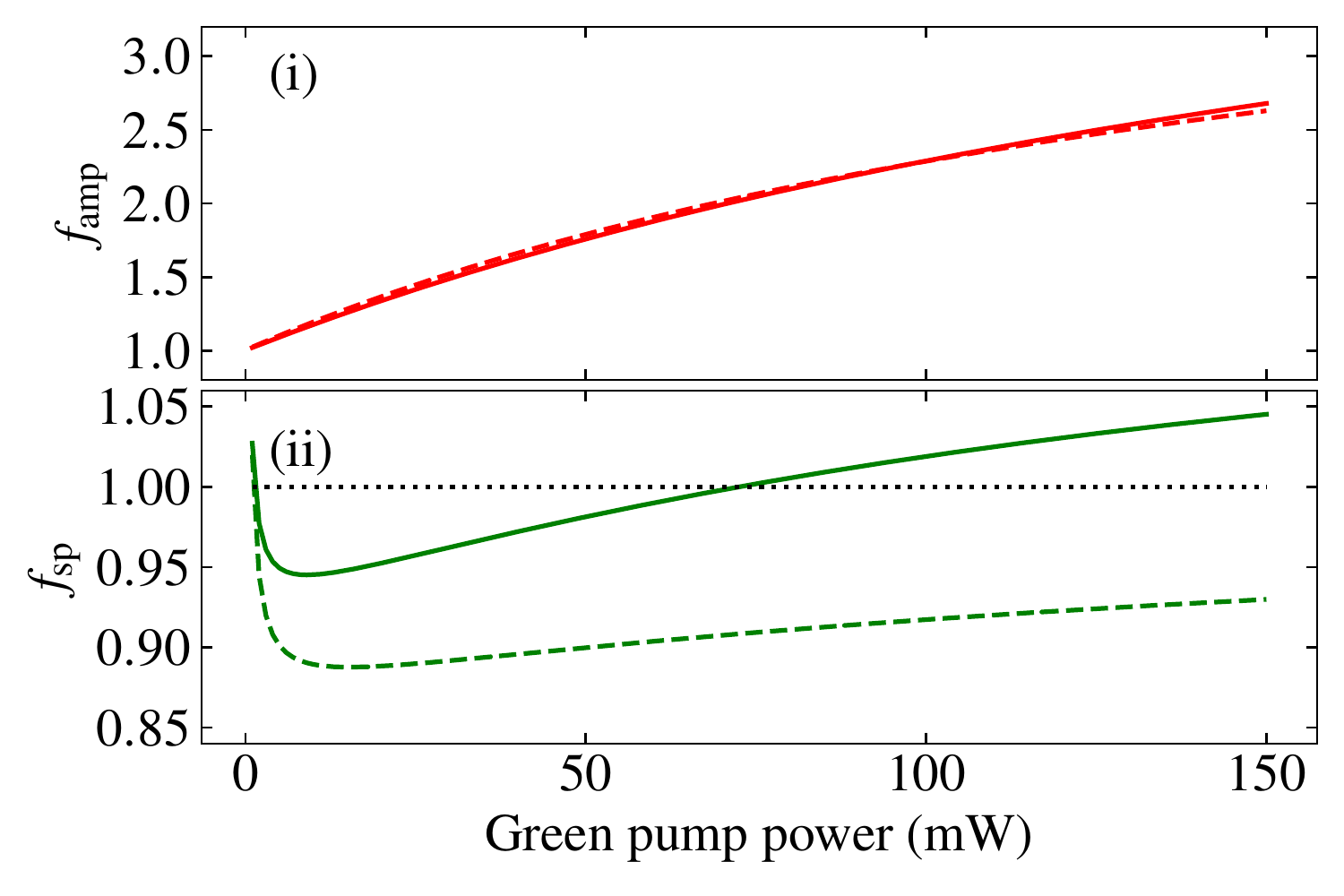}
\caption{$f_\mathrm{amp}$ and $f_\mathrm{sp}$ from the qualitative model as a function of green pump power. The solid curves corresponds to the model with charge state switching and the dashed curves corresponds to NV$^{-}$ centres without any charge state switching (\romannum{1}) The $f_\mathrm{amp}$ due to stimulated emission. (\romannum{2}) The $f_\mathrm{amp}$ spontaneous emission with the stimulated emission of both charge state of NV centre is shown in green solid line. The black dotted line is the reference line at factor 1.}
\label{fig:6}
\end{figure}

\begin{figure*}
\centering
\subfigure[]
{\label{fig:7.1}\includegraphics[scale = 0.5]{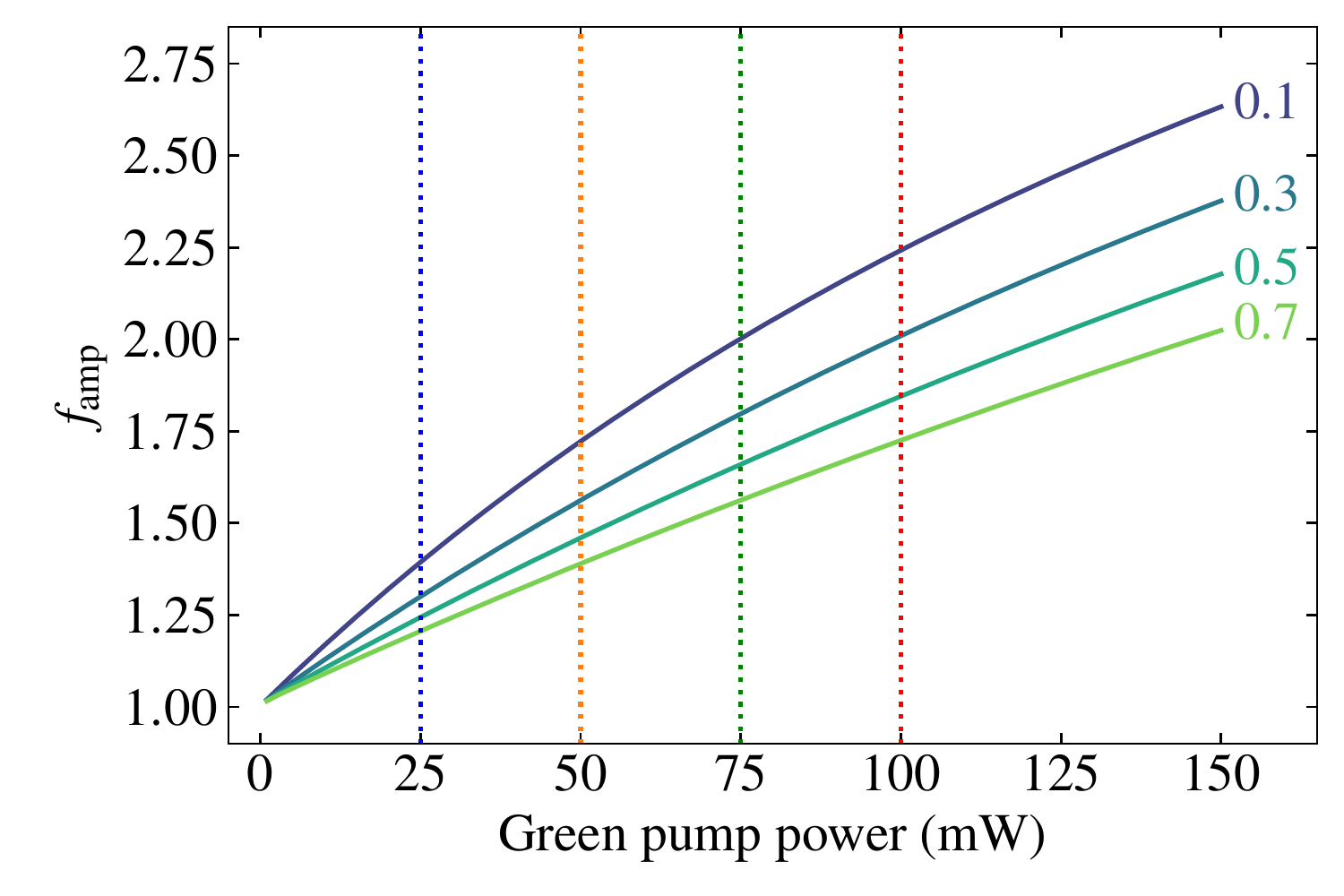}}
\quad\qquad
\subfigure[]
{\label{fig:7.2}\includegraphics[scale = 0.5]{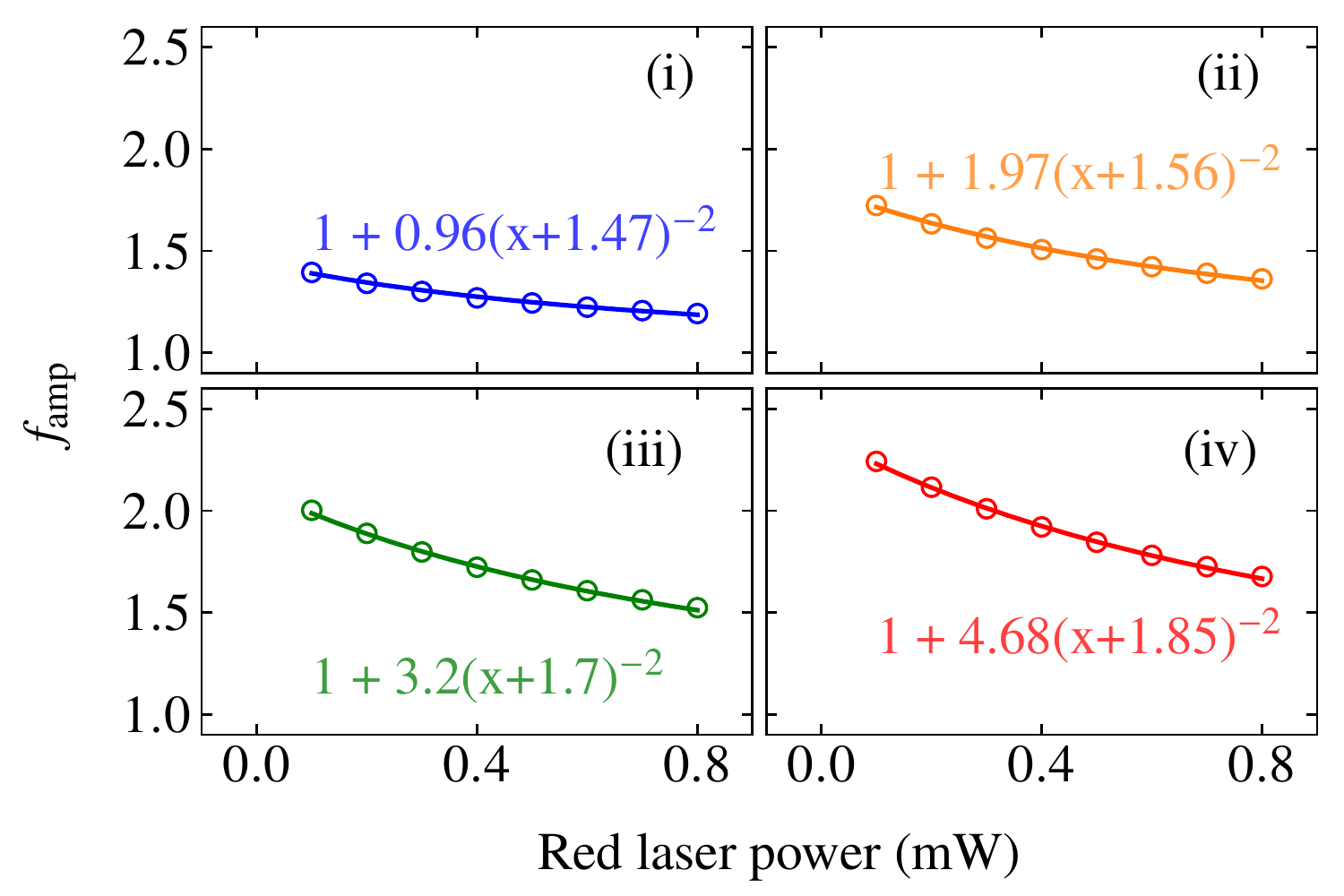}}
\caption{Change in amplification factors calculated from the theoretical model as a function of red laser power. The red laser power in mW is annotated for each cure in their respective colours. 
(b) The amplification factors in the dotted lines drawn in (a) as a function of red laser power. The blue, orange, green, and red circles in (\romannum{1}), (\romannum{2}), (\romannum{3}), and (\romannum{4}) are the amplification factors corresponding to the green pump powers 25 mW, 50 mW, 75 mW and 100 mW where the dotted lines drawn in (a) with respective colours. The solid curves with respective colors are the fit functions annotated in each figure.}
\label{fig:7}
\end{figure*}
\begin{figure*}
\centering
\subfigure[]
{\includegraphics[scale = 0.5]{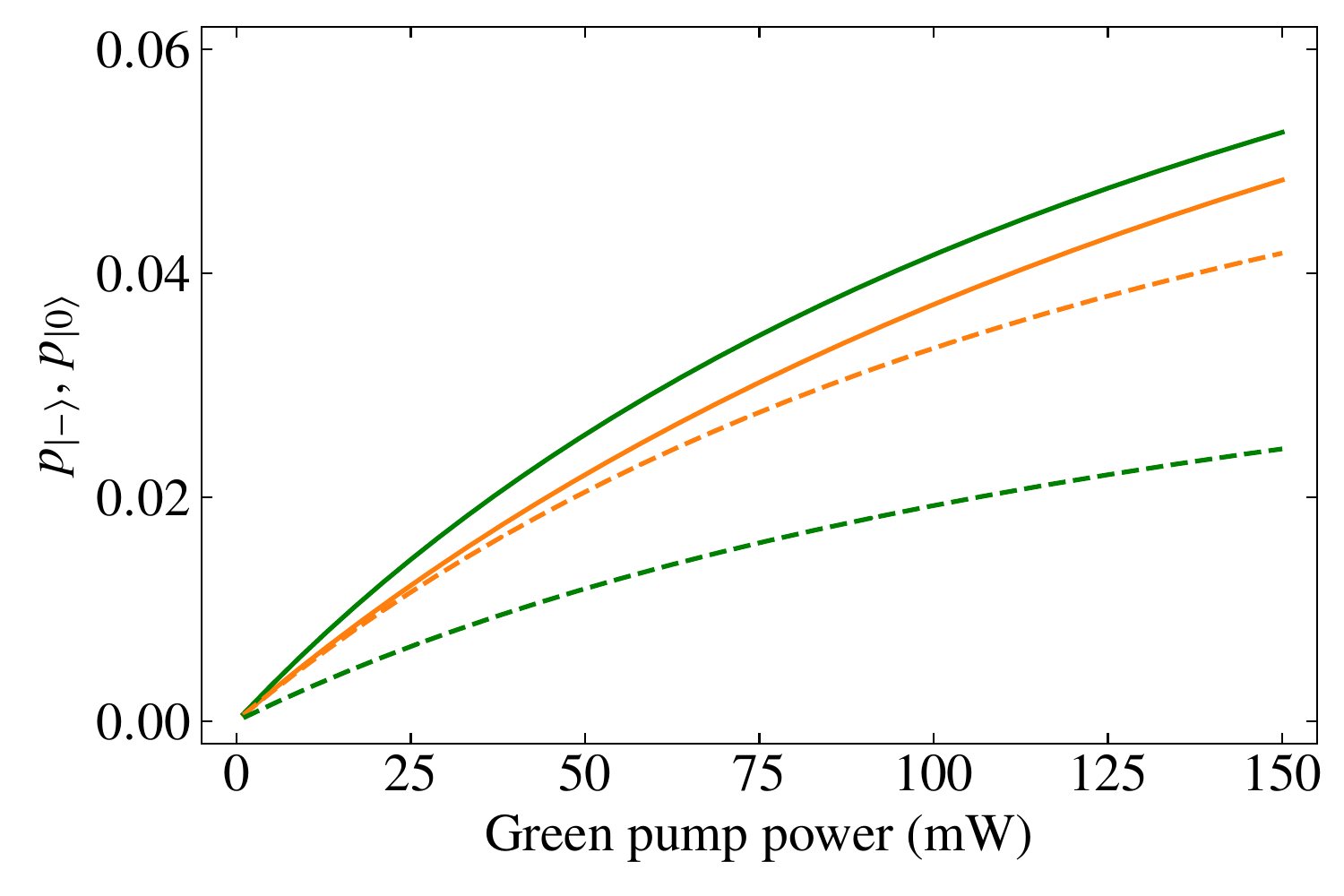}{\label{fig:8.1}}}
\quad\qquad
\subfigure[]
{\includegraphics[scale = 0.5]{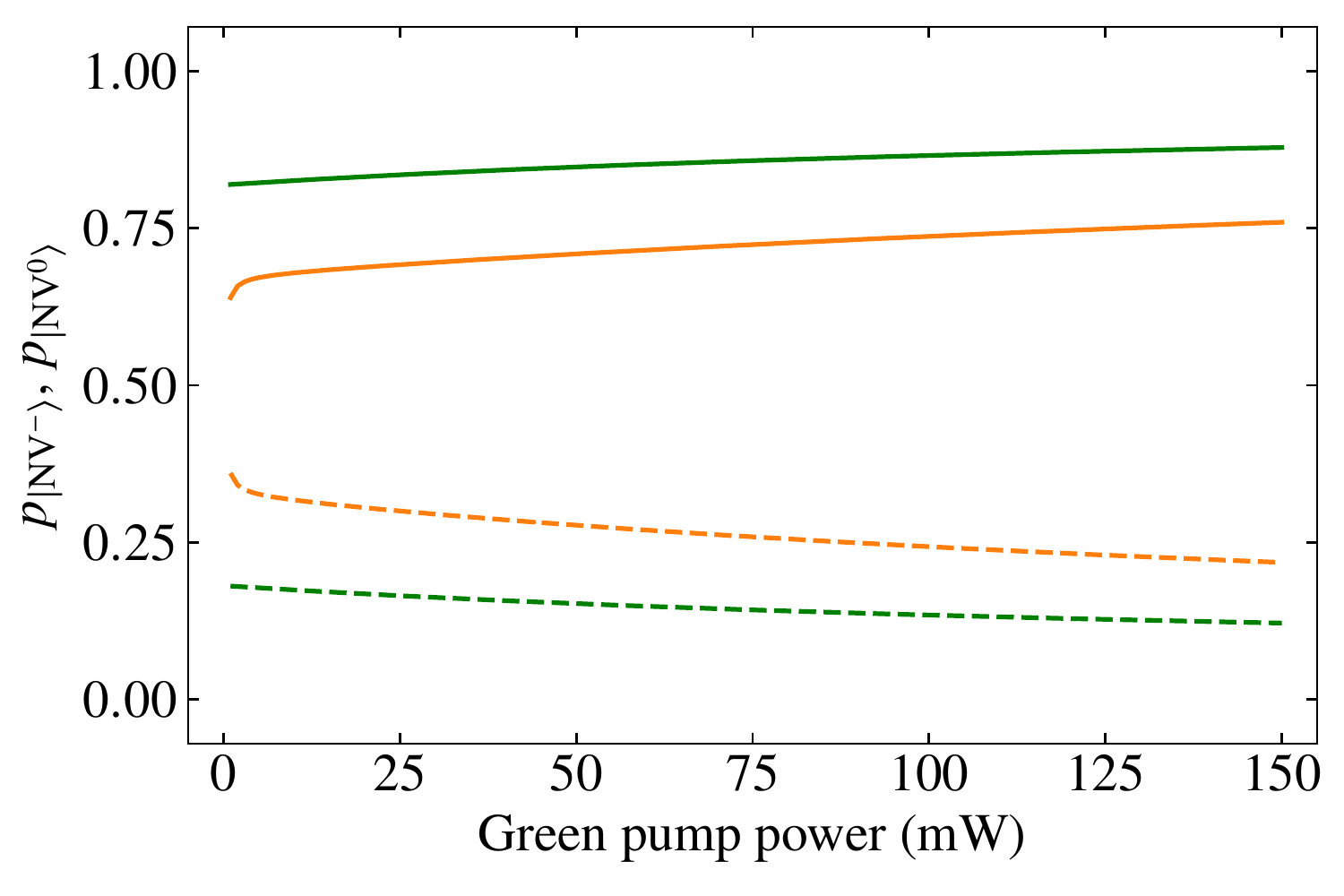}{\label{fig:8.2}}}
\caption{(a) $p_{\ket{-}}$ and $p_{\ket{0}}$ as a function green pump power. Green and orange colours represent green alone pumping and green plus red pumping respectively. 
The solid and dashed lines correspond to NV$^{-}$ and NV$^{0}$ respectively.
For green plus red pumping, stimulated emission of NV$^{-}$ and NV$^{0}$ is considered.
However, for green alone pumping, we neglect the possibility of stimulated emission from any of the charge states considering the fact that the spontaneous emission from these charge states into the cavity is not stimulating the emission as we do not see any trace of it in our experiment.
(b) Total fraction of  NV$^{-}$ and NV$^{0}$. The plot styles follows the same as (a).  
}
\label{fig:8}
\end{figure*}

$f_\mathrm{amp}$ and $f_\mathrm{sp}$ as defined in equations ~(\ref{eq:1}) and (\ref{eq:2}), respectively, are plotted as a function of green pump power in Figure \ref{fig:6}.
The exact parameters of the NV centres underlying the plots are discussed in the Appendix. 
We further assumed the only free parameter, $F$ $\sim$ 1200.
Given the design finesse of around 40000, the surface roughness and thickness of the diamond sample, this value seems reasonable. %
Both the calculated amplification factor $f_\mathrm{amp}$ and the spontaneous emission factor $f_\mathrm{sp}$ qualitatively match the ones observed in our experiments. 

To be completely forthright and transparent, whilst we have obtained values from diamond samples that are consistent with our own. A number of parameters and process can be modified whilst still capturing qualitatively similar results as our experiment.
There are two essential processes that must be occurring in order to match the trends we observed.
Firstly, the red 721~nm field must be inducing charge state switching preferencing ionisation over recombination, consistent with other experimental observations in the literature~\cite{chen_near-infrared-enhanced_2017,hacquebard2018charge,roberts2019spin}.
Secondly, in order to observe spontaneous emission factors, $f_\mathrm{sp}$ greater than one the NV$^0$ must either have a stronger excitation efficiency than the NV$^-$ under 532~nm illumination or be more efficiently collected. 
In our experiment the collection efficiency of each charge state, considering the spectral collection window, is approximately 1. 
Therefore, we must have a stronger excitation efficiency for NV$^0$ in our model which has been set to $\xi$ = 1.3 as consistent with Hacquebard \textit{et al.}~\cite{hacquebard2018charge} 
Unfortunately, the value of this second parameter is not consistent in the literature and ranges from 0.33 --- 1.8~\cite{hacquebard2018charge,roberts2019spin,manson_nv_2018,meirzada2018negative} and therefore requires further investigation to place concrete values and its sample dependency.

Our model is also able to qualitatively capture the observed dependence of the amplification factor $f_\mathrm{amp}$ on the red laser power (see Figure~\ref{fig:3}).
Figure \ref{fig:7} displays the calculated $f_\mathrm{amp}$ as a function of green pump power for different red pump powers, alongside the corresponding line cuts at constant green pump power, similar to the experimental curves in Figure~\ref{fig:3}. 
We observe a similar overall trend in $f_\mathrm{amp}$ as compared to experiments. 
In particular, we also observe the quadratic decrease of $f_\mathrm{amp}$ as function of red laser power for a fixed green pump power.
In order to understand the underlying physics, we also plot the calculated relevant populations of NV$^{-}$ and NV$^{0}$ in Figure~\ref{fig:8}.
This graph clearly reveals the influence of ionization on the observed cavity output.
At any green pump power, the population in NV$^{0}$ dominates over the NV$^{-}$ population.
The red laser enhances the ionization towards NV$^{0}$ as is evident from the fact that the orange dashed line representing the NV$^{0}$ population under green plus red pumping lies significantly higher than the green dashed line representing green pumping only.
Similarly, the solid orange line representing the NV$^{-}$ population under green plus red pumping lies below the solid green line for green pump only.
Furthermore, in our model, excited state population of each charge state that determines the gain is much smaller than total population of the respective charge state. This is due to the significant population in the ground state of both charge states and also in the singlet state for the NV$^{-}$ centre. 

\section{\label{sec:discussions}Discussion}
In the light of the model presented in the previous section, we interpret that the amplification at the \SI{721}{\nano\metre} observed in the experiment is due to the stimulated emission from both the charge states of NV centres.
In the experiments, due to the cavity-length tuning, the red seed wavelength can excite both fundamental Gaussian as well as higher-order Hermite-Gaussian spatial modes in the cavity.
Similarly, the emission of NV centres can be in the fundamental Gaussian as well as into the higher-order transverse modes of the cavity.
Since the coupling into the cavity mode and the spatial distribution of the cavity mode varies depending on the order of the mode~\cite{janitz2015fabry}, the amplification factor due to stimulated emission is also expected to have different values for the different cavity modes.
The maximum amplification is expected to happen in the fundamental mode of the cavity.
As a result the reduction in $f_\mathrm{amp}$ for the experiments compared with that of the theoretical estimation can be attributed to the effect of averaging of the amplification values due to the cavity-length tuning.

For the spontaneous emission dynamics in this model, photons emitted from the NV$^{0}$ charge state are playing a central role.
At low green pump powers the number of spontaneous photons is reduced when the red seed light is added.
Since the model without the ionization shows the same trend, the reduction in spontaneous emission can be attributed mainly to the stimulated emission.
As a result, $f_\mathrm{sp}$ reduces for small pump powers (evident both in the simulated and measured data).
As the green power increases, however, due to efficient pumping and a slightly longer excited-state lifetime, the NV$^{0}$ excited-state population, $p_{\ket{0}}$ begins to dominate over the NV$^-$ population, $p_{\ket{-}}$.
This implies that eventually spontaneous emission from NV$^{0}$ non-negligibly contribute to the fluorescence signal. 
Eventually, for higher green pump powers $f_\mathrm{sp}$ even increases above 1.
In the absence of ionization, the stimulated emission process should simply deplete the excited state and the maximum $f_\mathrm{sp}$ that one would expect is 1.
The simulated result for the model without ionization gives $f_\mathrm{sp}$, but increases for higher green pump powers.
We therefore interpret the observed characteristic trend in $f_\mathrm{sp}$ accompanying the measured $f_\mathrm{amp}$ for higher green pump powers comes from both the stimulated emission of NV$^{-}$ centres and charge state switching of these centres.

The minimum value for the $f_\mathrm{amp}$ for the model presented here is 1 and thus cannot account the negative amplifications that we mentioned in the experimental results.
Such cases can be attributed to  additional mechanisms such as excited state absorption and photo-thermal effects on the fibre etc.

\section{\label{sec:conclusion} Conclusion}

We have presented experimental data showing amplification of red laser light in an open fibre cavity filled with highly NV-doped diamond. 
The amplification increases with green pump power but there is a saturation effect and we do not obtain very strong absolute signals based on the amplification effect. 
Based on our qualitative model of the underlying NV colour center dynamics, we attribute the amplification to stimulated emission from NV centres. 
The observed effect of reduced spontaneous emission at lower pump powers and increased spontaneous emission at high powers, both on the order of a few per cent, was reproduced in our model by a photoionisation towards the NV$^0$. 
The observed effect thus arises from the fact that a fraction of the stimulated and spontaneous photons observed in the experiment (especially at higher green pump powers) are coming from the neutral charge state of the NV centre as well. 

The relative amplification factors indicate that considerable stimulated emission occurs in the cavity, however a strong absolute signal was not observed due to cavity loss channels other than the mirror transmissions, in particular due to the remaining absorption in the diamond material arising from either the \NV\,centres themselves or other internal defects. 
For the realisation of an NV laser in the future, this poses the challenges of reducing the remaining strong absorption due to other structural or nitrogen-related defects in highly NV-doped diamond, and highlights the importance of a stable charge state.
We conclude that a majority of the currently commercially available type 1b diamond shows absorption in the red, amplified through the multiple passes of a cavity prevents active standalone lasing in fibre-cavities.  
Having said that, much work remains to be done on exploring the physics and reducing the material loss channels, such as absorption and birefringece of highly NV-doped diamond for cavity and laser applications. 
This will involve substantial materials engineering including a detailed understanding of the nitrogen defects, structural defects, absorption from ground and excited states as well as charge donors and acceptors in the diamond crystal.

\begin{acknowledgements}
Sarath Raman Nair acknowledges that this work was done while holding International Macquarie University Research Excellence Scholarship (iMQRES) (allocation number 2014108) and also acknowledges EQUS centre collaboration award. Lachlan Rogers is the recipient of an Australian Research Council Discovery Early Career Award (project number DE170101371). This work was supported by Australian Research Council centre of Excellence for Engineered Quantum Systems (CE170100009). Jan Jeske acknowledges funding from the German federal ministry for education and research Bundesministerium f\"ur Bildung und Forschung (BMBF) under grant number  13XP5063. Andrew D. Greentree acknowledges Australian Research Council under the Future Fellowship (FT160100357).
\end{acknowledgements}

\section*{contributions}
JJ, ADG, TV, XV, SR. conceived the initial ideas about the experiments.
JJ, XV, SR, LJR, TV planned the experiments.
HO, TO, TY and JJ prepared and processed the diamond sample. 

SR and LJR built the experiments with contributions from XV.
SR took measurements and analysed the data with contributions from LJR.
SR performed the theoretical modelling and simulations.
SR, TV, LJR, RPR, and JJ interpreted the results.
SR, TV, LJR and RPR wrote the manuscript with varying contributions from all authors.
TV supervised the project.

\appendix
\section{\label{sec:addendixSwitching}Charge state switching dynamics}

\begin{figure*}
\centering
\includegraphics[scale = 0.5]{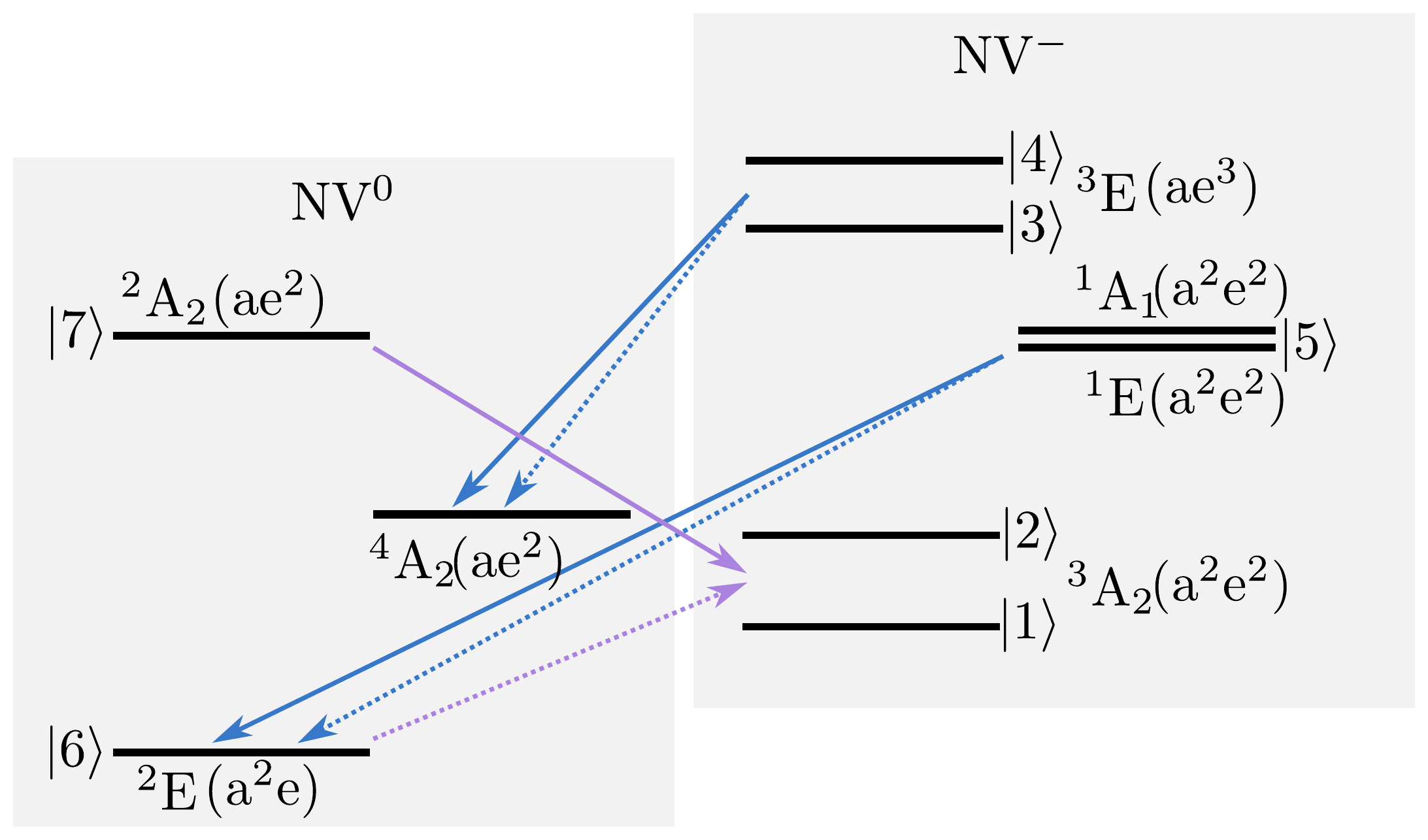}
\caption{Electronic energy level scheme of the NV centre showing the ionisation and recombination processes that have been observed. 
Levels 1 to 5 belong to NV$^{-}$ and 6 and 7 belong to NV$^{0}$ and have been labeled with their corresponding molecular symmetry assignments as well as their molecular electron orbital states. This picture shows that state 5 is in fact two distinct energy levels and there is also a spin quartet meta-stable state that plays a role in the ionisation dynamics. These states have been removed in Figure~\ref{fig:4} as they do not significantly impact the rate equation description of the system.
The blue arrows indicate ionisation pathways from the NV$^-$ state to the NV$^0$ state, whereas the purple arrows indicate recombination pathways from NV$^0$ to NV$^-$. 
The dashed arrows indicate tunneling processes that are mediated by a nearby single substitutional nitrogen acting as a charge donor to the NV centre. The solid lines are photon mediated processes. 
At this stage it is uncertain if all of these photon mediated processes are intrinsic to the NV centre or are due to ionisation of nearby charge donors leading to a tunneling process.
It is unlikely that ionisation from the $^3$E(ae$^3$) excited state of the NV$^-$ occurs directly to the $^2$E(a$^2$e) ground state of the NV$^0$ as this must involve a two electron process. 
It is more likely that the transition occurs into the $^4$A$_2$(ae$^2$) meta-stable state of the NV$^0$ centre before quickly and non-radiativly decaying into the ground state of the NV$^0$.
In the rate equation models used to simulate this system these two processes are approximately equivalent.
It is important to note that there may be additional allowed pathways for ionisation such as from the $^3$E(ae$^3$) excited state of the NV$^-$ into the $^2$A$_2$(ae$^2$) excited state of the NV$^0$ that we have not considered in this paper. 
}
\label{fig:A1}
\end{figure*}
To convert between the two charge states we need to examine both the ionisation process from NV$^-$ to NV$^0$ and the recombination process from NV$^0$ to NV$^-$ as depicted in Figure~\ref{fig:A1}. 
The charge state switching dynamics in NV centres is a complex problem that is still being studied actively in the community. 
One problem is that the dynamics depend strongly on the properties of the diamond; in particular, the density of the NV centres, the surrounding singlet nitrogen charge donors and the impact of surface charge acceptors all play a significant role, varying dramatically between diamond samples. 
Including the properties of the singlet nitrogen in the diamond lattice is essential for understanding the dynamics as they are the main defect responsible for supplying the NV centres with an additional charges~\cite{collins_fermi_2002}.
In the dark, the charge state of individual NV centres depends on the separation distance between the NV centre and the singlet nitrogen. 
For small separations  ($\sim$5~nm) the NV--N combination an electron will tunnel from the singlet nitrogen to the NV centre to produce an NV$^-$ -- N$^+$ pair~\cite{manson_nv_2018}. 
For these diamonds the conversion of the NV$^-$ centre into the NV$^0$ charge state can occur in via single photon process. 
The NV$^-$ centre must first be excited into the excited state where the electron can then tunnel back to the singlet nitrogen leaving the NV centre converted into the NV$^0$ charge state.
As mentioned by Manson \textit{et al.} tunneling from the excited state of the NV$^-$, $^3$E(e$^3$a) state, to the ground state of the NV$^0$, $^2$E(a$^2$e) is unlikely to occur as this would involve a two-electron transition.
It is therefore more likely that the NV$^-$ excited state, $^3$E(e$^3$a) tunnels directly to the meta-stable $^4$A$_2$(e$^2$a) quartet level before decaying non radiatively to the NV$^0$ ground state, $^2$E(a$^2$e). 
Additionally it is also possible that tunneling occurs through the NV$^-$ meta-stable, $^1$E$_1$(a$^2$e$^2$) state to the ground state of NV$^0$, $^2$E(a$^2$e). 
It is clear that the specific details of the tunneling transitions require further experimental and theoretical consideration.
For larger NV--N separations as well as for higher intensity excitation, ionisation from NV$^-$ to NV$^0$ occurs through a two-photon process. First the NV$^-$ centre must be excited and then subsequently a second electron is stripped from the NV$^-$ complex.
This second ionisation step occurs either from the excited $^3$E(e$^3$a) state or from the $^1$E$_1$(a$^2$e$^2$) state of the NV$^-$~\cite{hacquebard2018charge}. Using similar arguments as above it is likely that this will produce NV$^0$ centres in the meta-stable $^4$A$_2$(e$^2$a) quartet level, or ground $^2$E(a$^2$e) state respectively. 
Interestingly, both the 532~nm excitation laser and 721~nm seed laser can induce both of these processes, the 721~nm seed laser only weakly excites the NV$^-$ centre and more efficiently induces the second ionisation step. The result of this is that increasing the 721~nm field intensity leads to a stronger non-radiative transition to the NV$^0$ ground state.
In addition, using 721~nm CW excitation alone, negligible fluorescence will be observed since the excitation cross section is small and this laser preferentially populates the NV$^0$ charge state. The NV$^0$ charge state can then not be excited at this wavelength resulting in no fluorescence signal.

The recombination process from NV$^0$ to NV$^-$ can also occur through a tunneling process or a two-photon process.
The tunneling process for close NV--N pairs as described above is caused by a singlet nitrogen providing the electron which is captured by the NV$^{0}$ in its $^2$E(e$^2$a) ground state tunneling to the NV$^-$ in its $^3$A$_2$(a$^{2}$e$^{2}$) ground state. 
The two-photon process from NV$^0$ to NV$^-$ occurs again by accepting a charge from a nearby charge donor. 
Firstly, a photon must excite the NV$^0$ into its $^2A$(ae$^2$) excited state. Subsequently, a second electron can then be excited from the valence band, providing the additional electron required to turn transition the centre into the NV$^-$, $^3$A$_2$(a$^2$e$^2$) ground state~\cite{siyushev_optically_2013}.
The second step of this process is considered to be mediated by charge donors in the diamond lattice.
For typical diamonds that contain NV centres this is dominated by the surrounding singlet nitrogen which can be ionised for energies from 1.7---2.2~eV~\cite{enckevort_temperature_1992,davies_vacancy-related_1992}. 
At the lower ionisation estimates, both the 532~nm and 721~nm fields can ionise the singlet substitutional nitrogen leading to extra free electrons producing NV$^-$ centres. 
Interestingly, an increased the recombination rate has been observed for under NIR illumination for ensembles of NVs in nano-diamonds at 785~nm~\cite{roberts2019spin} and for single NV centres in bulk at 766~nm~\cite{hacquebard2018charge}. 
Both of these wavelengths should not have enough energy to excite the singlet nitrogen, nevertheless the recombination is still observed. 
For ensembles the increase in red induced recombination could be arising from an increased rate of ionisation of the surrounding NV$^-$ centres in the diamond, producing free electrons for NV$^0$ recombination. 
However, for single NV centres this effect should not be possible.
This wavelength independancy of the recombination raises questions over the exact processes involved in the charge state switching. 
It is possible therefore that the ionisation and recombination processes of the NV centres are not intrinsic to the NV centres but mediated by the broadband ionisation of donors and acceptors within the diamond lattice causing NV tunneling rates to increase.
If this is indeed the case then it may still be possible to engineer diamond samples with stabilised NV$^-$ centres that are less sensitive to the intense NIR fields required for some applications including the NV laser magnetometer.

\section{\label{sec:appendix2}Stimulated emission cross-section}

\begin{figure}
\centering
\includegraphics[scale = 0.5]{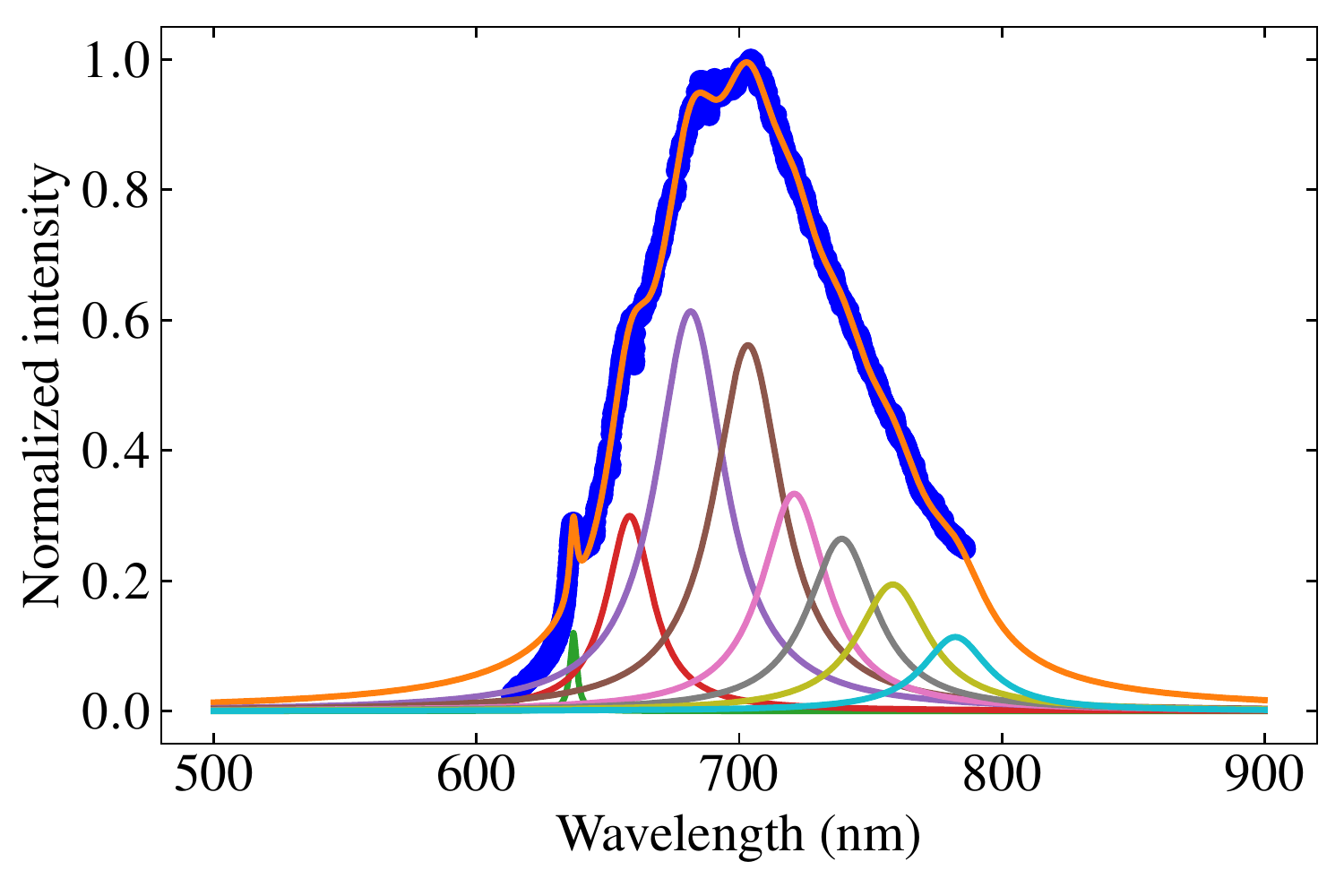}
\caption{Spectrum of the NV centre from our sample by green wavelength pumping without any cavity mirrors.
The blue dots are the experimentally obtained spectrum. The orange curve is the fit using a convolution of 8 Lorentzian functions for the Zero-Phonon line (ZPL) and 7 phonon levels.
The individual Lorentzian functions are also shown. See the text for the details.}
\label{fig:A2}
\end{figure}

We estimate the stimulated emission of the NV centre in a similar way of \cite{fraczek2017laser}, but without specification to the charge state.
For this we obtain the spectrum of the NV centre normalized to the peak intensity, without any cavity for an arbitrary green pump power.
This normalized spectrum is shown in Figure \ref{fig:A2}.
The fit parameters for the Lorentzian functions obtained from the fit is shown in Table~\ref{tab:2}.
The fit shows that the peak position of one of the phonon levels is $\sim$ 721 nm, at the red wavelength.
However, since the phonon levels is expected to have extremely short life time, that induces very strong dephasing, we neglect any coherent transition from the phonon levels similar to the stimulated absorption case.

\begin{table}
\centering
\begin{tabular}{llc}
\hline
\textbf{Peak position}  & \textbf{amplitude} & \textbf{FWHM}\\
\hline
 636.9 nm & 0.1 & 3 nm \\
658.3 nm & 0.3 & 21 nm\\
681.5 nm & 0.6 & 31.7 nm\\
703.4 nm & 0.6 & 31.7 nm\\
721.0 nm & 0.3 & 30.4 nm\\
739.0 nm & 0.3 & 31.7 nm\\
758.5 nm & 0.2 & 33.1 nm\\
782.3 nm & 0.1 & 30.4 nm\\
\hline
\end{tabular}
\caption{Parameters obtained from the 8 Lorentian fit functions of the spectrum.}
\label{tab:2}
\end{table}

\begin{figure}
\centering
\includegraphics[scale = 0.5]{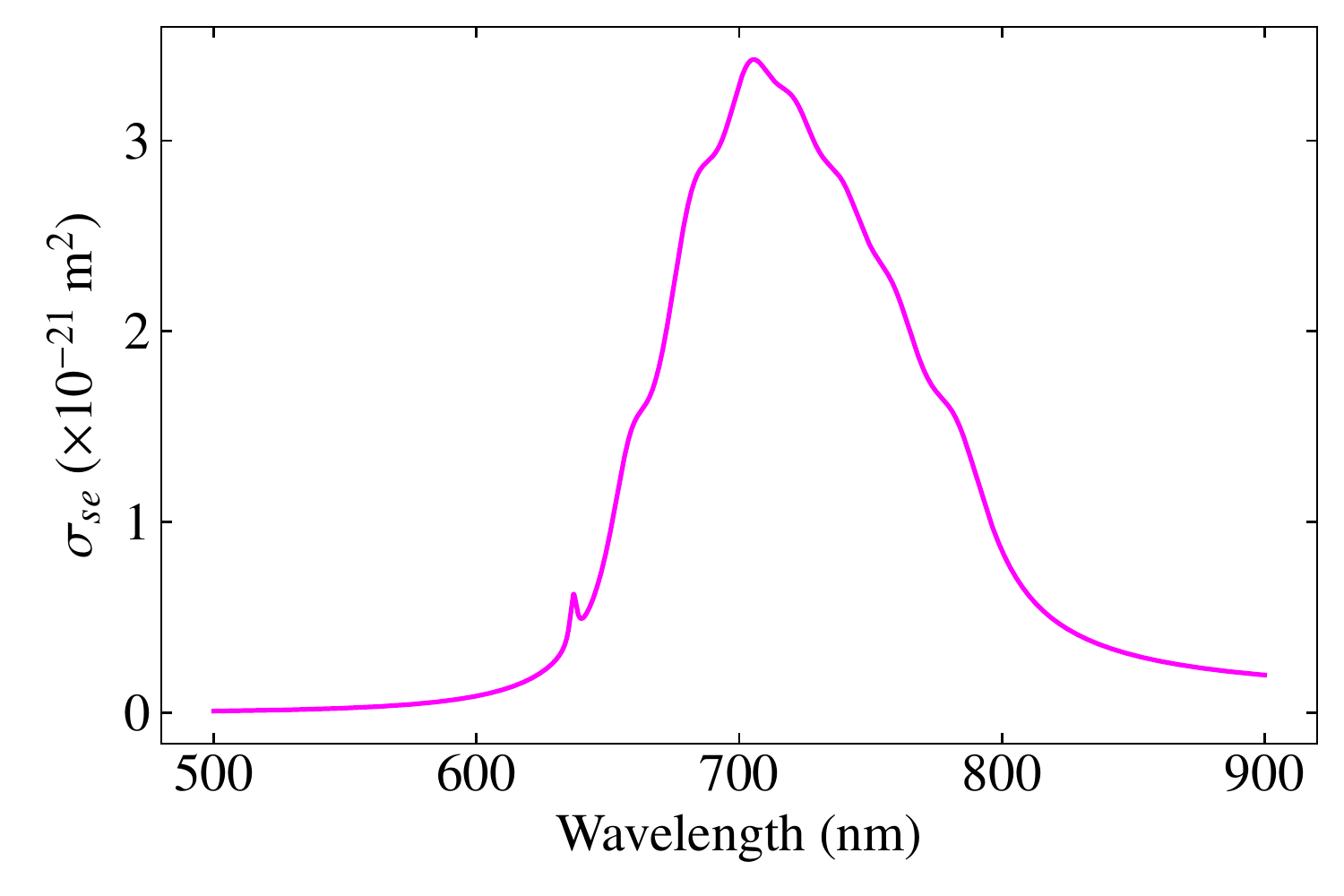}
\caption{Spectrum of the NV centre from our sample by green wavelength pumping without any cavity mirrors.
Stimulated emission of our sample as a function wavelength. At $\sim$ 721nm, $\sigma_{se} \sim 3.22 \times 10^{-21}$ m$^{-2}$.}
\label{fig:A3}
\end{figure}

We calculate the stimulated emission cross-section using the F\"uchtbauer-Ladenburg equation given as \cite{trager2012springer, aull1982vibronic, fraczek2017laser},
\begin{equation}
\label{eq:10}
\sigma_{se}(\lambda) = \frac{1}{8\pi n^{2} c \gamma}\frac{\lambda^{5}I(\lambda)}{\int{\lambda I(\lambda)}d\lambda}.
\end{equation}
Here $n$ is the refractive index of the diamond sample, $c$ speed of light, $\gamma$ is the spontaneous decay rate, $\lambda$ is the wavelength and $I(\lambda)$ is the intensity as a function of wavelength.
We replace $\int{\lambda I(\lambda)}d\lambda$ in equation (\ref{eq:10}) with $\sum{\lambda I(\lambda)}\delta \lambda$.
Then we obtain $\sigma_{se}(\lambda)$ from the normalized spectrum (Figure (\ref{fig:A2})) and the result is shown in Figure \ref{fig:A3}.
To find the stimulated emission cross-section, without specification to the charge state, we consider $\gamma$ as $r_{31}$.

In our experiment we do not observe any saturation of the spontaneous emission from the NV centres.
Hence we consider a lower value of the absorption cross-section for the green available in literature from references \cite{dumeige2019infrared, wee2007two}.
Recent results on nano-diamond NV ensemble~\cite{roberts2019spin} suggests that ionization rates are spin dependent when the NV centres are pumped with green and red wavelengths.
We do not directly include spin dependant ionisation from the excited state but include an indirect spin dependence on the ionisation through the spin-dependent transition to the singlet states which can then be ionised~\cite{hrubesch2017efficient}.
We set the $\sigma^s_{I} = 0.0215 \times \sigma_{se}$ as consistent with Hacquebard~\textit{et al.}~\cite{hacquebard2018charge}.

\bibliography{apssamp}

\end{document}